\documentclass[12pt]{article}
\usepackage{amssymb}
\usepackage{amsmath}
\usepackage{apacite}
\usepackage{latexsym}
\usepackage{epsfig}
\usepackage{graphicx}
\usepackage{subfigure}
\usepackage{wasysym}
\usepackage{threeparttable}
\usepackage{natbib}
\usepackage{color}
\usepackage{epstopdf}
\usepackage{bm}
\usepackage{float}
\usepackage{todonotes}
\usepackage{verbatim}
\usepackage{booktabs}

\usepackage{hyperref}

\def\log{\hbox{log}}

\def\boxit#1{\vbox{\hrule\hbox{\vrule\kern6pt
          \vbox{\kern6pt#1\kern6pt}\kern6pt\vrule}\hrule}}

\def\bse{\begin{eqnarray*}}
\def\ese{\end{eqnarray*}}
\def\be{\begin{eqnarray}}
\def\ee{\end{eqnarray}}
\def\bq{\begin{equation}}
\def\eq{\end{equation}}
\def\bse{\begin{eqnarray*}}
\def\ese{\end{eqnarray*}}

\begin{document}

\thispagestyle{empty} 
\baselineskip=28pt

\begin{center}
{\LARGE{\bf Unit Level Modeling of Survey Data for Small Area Estimation Under
    Informative Sampling: A Comprehensive Overview with Extensions}}

\end{center}

\baselineskip=12pt

\vskip 2mm
\begin{center}
Paul A. Parker\footnote{(\baselineskip=10pt to whom correspondence should be addressed)
Department of Statistics, University of Missouri,
146 Middlebush Hall, Columbia, MO 65211-6100, paulparker@mail.missouri.edu},

 Ryan Janicki\footnote{\baselineskip=10pt Center for Statistical Research and
  Methodology, U.S. Census Bureau, 4600 Silver Hill Road, Washington,
  D.C. 20233-9100, ryan.janicki@census.gov},

   and Scott H. Holan\footnote{\baselineskip=10pt Department of Statistics, University of Missouri,
146 Middlebush Hall, Columbia, MO 65211-6100, holans@missouri.edu}\,\footnote{\baselineskip=10pt Office of the Associate Director for Research and Methodology, U.S. Census Bureau, 4600 Silver
  Hill Road, Washington, D.C. 20233-9100, scott.holan@census.gov}

\end{center}

\vskip 4mm
\baselineskip=12pt 
\begin{center}
{\bf Abstract}
\end{center}

Model-based small area estimation is frequently used in conjunction with survey data
in order to establish estimates for under-sampled or unsampled geographies.  These
models can be specified at either the area-level, or the unit-level, but unit-level
models often offer potential advantages such as more precise estimates and easy
spatial aggregation. Nevertheless, relative to area-level models, literature on
unit-level models is less prevalent. In modeling small areas at the unit level,
challenges often arise as a consequence of the informative sampling mechanism used to
collect the survey data. This paper provides a comprehensive methodological review
for unit-level models under informative sampling, with an emphasis on Bayesian
approaches. To provide insight into the differences between methods, we conduct a
simulation study that compares several of the described approaches. In addition, the
methods used for simulation are further illustrated through an application to the
American Community Survey. Finally, we present several extensions and areas for
future research.

\baselineskip=12pt
\par\vfill\noindent
\par\medskip\noindent
\clearpage\pagebreak\newpage \pagenumbering{arabic}
\baselineskip=24pt


\section{Introduction}

Government agencies have seen an increase in demand for data products in recent
years. One trend that has accompanied this demand is the need for granular estimates
of parameters of interest at small spatial scales or subdomains of the finite
population.  Typically, sample surveys are designed to provide reliable estimates of
the parameters of interest for large domains. However, for some subpopulations, the
area-specific sample size may be too small to produce estimates with adequate
precision.  The term {\em small area} is used to refer to any domain of interest,
such as a geographic area or demographic cross classification, for which the
domain-specific sample size is not large enough for reliable direct estimation.  To
improve precision, model-based methods can be used to `borrow strength,' by relating
the different areas of interest through use of linking models, and by introducing
area-specific random effects and covariates.  The Small Area Income and Poverty
Estimates (SAIPE) program, and the Small Area Health Insurance Estimates (SAHIE)
program within the U. S. Census Bureau are two examples of government programs which
produce county and sub-county level estimates for different demographic cross
classifications across the entire United States using small area estimation (SAE)
methods \citep{luery11, bau18}. It can be difficult to generate small area estimates
such as these for a number of reasons, including the fact that many geographic areas
may have very limited sample sizes, if they have been sampled at all.


Models for SAE may be specified either at the area level or the unit level
\citep[see][for an overview of small area estimation methodology]{rao15}. Area-level
models treat the direct estimate (for example, the survey-weighted estimate of a
mean) as the response, and typically induce some type of smoothing across areas. In
this way, the areas with limited sample sizes may ``borrow strength" from areas with
larger samples.  While area-level models are popular, they are limited, in that it is
difficult to make estimates and predictions at a geographic or demographic level that
is finer than the level of the aggregated direct estimate.

In contrast, unit-level models use individual survey units as the response data,
rather than the direct estimates.  Use of unit-level models can overcome some of the
limitations of area-level models, as they constitute a bottom-up approach (i.e., they
utilize the finest scale of resolution of the data).  Since model inputs are at the
unit-level (person-level, household-level, or establishment-level), predictions and
estimates can be made at the same unit-level, or aggregated up to any desired level.
Unit-level modeling also has the added benefit of ensuring logical consistency of
estimates at different geographic levels.  For example, model-based county estimates
are forced to aggregate to the corresponding state-level estimates, eliminating the
need for ad hoc benchmarking.  In addition, because the full unit-level data set is
used in the modeling, rather than the summary statistics used with area-level models,
there is potential for improved precision of estimated quantities.

Although unit-level models may lead to more precise estimates, that aggregate
naturally across different spatial resolutions, they also introduce new
challenges. Perhaps the biggest challenge is how to account for the survey design in
the model.  With area-level models, the survey design is incorporated into the model
through specification of a sampling distribution (typically taken to be Gaussian) and
inclusion of direct variance estimates.  With unit-level models, accounting for the
survey design is not as simple.  One challenge is that the sample unit response may
be dependent on the probability of selection, even after conditioning on the design
variables. When the response variables are correlated with the sample selection
variables, the sampling scheme is said to be {\em informative}, and in these
scenarios, in order to avoid bias, it is critical to capture the sample design in the model by including the survey weights or the design variables used to construct the survey weights.



The aim of this paper is to present a comprehensive literature review of unit-level
small area modeling strategies, with an emphasis on Bayesian approaches, and to
evaluate a selection of these strategies by fitting different unit-level models on
both simulated data, and on real American Community Survey (ACS) micro-data, thereby
comparing model-based predictions and uncertainty estimates.  In this paper we focus
mainly on model specification and methods which incorporate informative sampling
designs into the small area model.  Some important, related issues, that will be
outside the scope of this paper include issues related to measurement error and
adjustments for nonresponse.  Generally, we assume that observed survey weights have
been modified to take into account nonresponse.  We also avoid discussion on the
relative merits of frequentist versus Bayesian methods for inference.  In the
simulation studies and data examples given in Sections~\ref{sec: sim} and \ref{sec:
  DA}, we fit three unit-level small area Bayesian models, with vague, proper priors
on all unknown model parameters.  Inference on the finite population parameters of
interest is done using the posterior mean as a point estimate, and the posterior
variance as a measure of uncertainty.




Some related work includes \citet{hid16}, who present a simulation study to compare
area-level and unit-level models, both fit in a frequentist setting. They fit their
models under both informative and noninformative sampling, and found that overall the
unit-level models lead to better interval coverage and more precise
estimates. \citet{gel07} discusses poststratification using survey data, and compares
the implied weights of various models including hierarchical
regression. \citet{lum17} discuss general techniques for modeling survey data with
included weights. They focus on frequentist pseudo-likelihood estimation as well as
hypothesis testing.   Chapter 7 of \citet{rao15} provides an overview of some commonly used
unit-level small area models.  The current paper adds to this literature by providing
a comprehensive review of unit-level small area modeling techniques, with a focus on
methods which account for informative sampling designs.  We mainly use Bayesian
methods for inference, but note that many model-based methods are general enough to
be implemented in either setting, and we highlight some scenarios where Bayesian
methodology may be used.



The remainder of this paper is organized as follows.  Section~\ref{sec: notation}
introduces the sampling framework and notation to be used throughout the paper.  We
aim to keep the notation internally consistent.  This may lead to differences
compared to the original authors' notation styles, but should lead to easier
comparison across methodologies.  In Section~\ref{sec: uw} we cover modeling
techniques that assume a noninformative survey design.  The basic unit-level model is
introduced, as well as extensions of this model which incorporate the design
variables and survey weights.  Methods which allow for an informative design are then
discussed, beginning in Section~\ref{sec: pseudo}.   Here, we discuss
  analytic inference of population parameters under an informative design using
  pseudo-likelihood methods.  Extensions of the pseudo-likelihood to hierarchical,
  multilevel mixed models are discussed, as well as application to small area
  estimation problems.  In Section~\ref{sec: samp_dist} we focus on models that
use a sample distribution that differs from the population distribution. We conclude
the review component of this paper in Section~\ref{sec: bin}, where we will review
models that are specific to a Binomial likelihood, as many variables collected from
survey data are binary in nature . In Section~\ref{sec: sim} we compare three
selected models to a direct estimator under a simulation study designed around
American Community Survey (ACS) data. Specifically, this simulation examines three
Bayesian methods that span different general modeling approaches (pseudo-likelihood,
nonparametric regression on the weights, and differing sample/population likelihoods)
with the goal of examining the utility of each approach. The Stan code used to fit these models is available at \url{https://github.com/paparker/Unit_Level_Models}. Similarly, Section~\ref{sec:
  DA} uses the same models for a poverty estimates application similar to the Small
Area Income and Poverty Estimates program (SAIPE). Finally, we provide concluding
remarks in Section~\ref{sec: conc}.


\section{Background and notation}\label{sec: notation}

Consider a finite population \(\mathcal{U}\) of size \(N\), which is subset into
\(m\) nonoverlapping domains, \(\mathcal{U}_i = \left\{ 1, \dots, N_i \right\}, i =
1, \dots, m\), where \(\sum^m_{i = 1} N_i = N\).  These subgroups will typically be
small areas of interest, or socio-demographic cross-classifications, such as age by
race by gender within the different counties. We use \(y_{ij}\) to represent a
particular response characteristic associated with unit \(j \in \mathcal{U}_i\), and
\(\boldsymbol{x}_{ij}\) a vector of predictors for the response variables \(y_{ij}\).

Let \(\boldsymbol{Z}\) be a vector of design variables which characterize the
sampling process.  For example, \(\boldsymbol{Z}\) may contain geographic variables
used for stratifying the population, or size variables used in a probability
proportional to size sampling scheme.  A sample \(\mathcal{S} \subset \mathcal{U} =
\bigcup \mathcal{U}_i\) is selected according to a known sampling design with
inclusion probabilities dependent on the design variables, \(\boldsymbol{Z}\).  Let \(\mathcal{S}_i\) denote the sampled units in small area \(i\), and
let \(\pi_{ij} = P(j \in \mathcal{S}_i \mid \boldsymbol{Z})\).  The inverse
probability sampling weights are denoted with \(w_{ij} = 1 / \pi_{ij}\).  We note
that as analysts, we may not have access to the functional form of \(P \left( j \in
\mathcal{S}_i \mid \boldsymbol{Z} \right)\), and may not even have access to the design
variables \(\boldsymbol{Z}\), so that the only information available to us about the
survey design is through the observed values of \(\pi_{ij}\) or \(w_{ij}\), for the
sampled units in the population.  Finally, we let \(D_S = \left\{ \left\{ y_{ij},
\boldsymbol{x}_{ij}, w_{ij} \right\}: j \in \mathcal{S}_i, i = 1, \dots, m \right\}\)
represent the observed data. This simply consists of the responses, predictors, and
sampling weights for all units included in the sample.  In this context,
$y_{ij}$ is random, $x_{ij}$ is typically considered fixed and known, and $w_{ij}$
can either be fixed or random depending on the specific modeling assumptions. 

The usual inferential goal, and the main focus of this paper, is on estimation of the
small area means, \(\bar{y}_i = \sum_{j \in \mathcal{U}_i} y_{ij} / N_i\), or totals,
\(y_i = \sum_{j \in \mathcal{U}_i} y_{ij}\).  In some situations, interest could be
on estimation of descriptive population parameters, such as regression coefficients
in a linear model, or on estimation of a distribution function.  The best predictor,
\(\hat{\bar{y}}_i\) of \(\bar{y}_i\), under squared error loss, given the observed
data \(D_S\), is
\begin{equation}\label{E: bp}
  \hat{\bar{y}}_i = E \left( \bar{y}_i \mid D_S \right) = \frac{1}{N_i} \sum_{j
    \in \mathcal{U}_i} E \left( y_{ij} \mid D_S \right) = \frac{1}{N_i} \sum_{j \in
    \mathcal{S}_i} y_{ij} + \frac{1}{N_i} \sum_{j \in \mathcal{S}^c_i} E \left(
    y_{ij} \mid D_S \right).
\end{equation}
The first term on the right hand side of \eqref{E: bp} is known from the observed
sample.  However, computation of the conditional expectation in the second term
requires specification of a model, and potentially, depending on the model specified,
auxiliary information, such as knowledege of the covariates \(\boldsymbol{x}_{ij}\)
or sampling weights \(w_{ij}\) for the nonsampled units.  
For the case where the predictors \(\boldsymbol{x}_{ij}\) are categorical, the
assumption of known covariates for the nonsampled units is not necessarily
restrictive, if the totals, \(N_{i, g}\), for each cross-classification \(g\) in each
of the small areas \(i\) are known.  In this case, the last term in \eqref{E: bp}
reduces to \(N^{-1}_i \sum_g (N_{i, g} - n_{i, g}) E \left( y_{ij} \mid D_S
\right)\), and only predictions for each cross-classification need to be made.

The predictor given in \eqref{E: bp} is general, and the different unit-level
modeling methods discussed in this paper are essentially different methods for
predicting the nonsampled units under different sets of assumptions on the finite
population and the sampling scheme.  An entire finite population can then be
generated, consisting of the observed, sampled values, along with model-based
predictions for the nonsampled individuals.  The small area mean can then be
estimated by simply averaging appropriately over this population.  If the sampling
fraction \(n_i / N_i\) in each small area is small, inference using predicted values
for the entire population will be nearly the same as inference using a finite
population consisting of the observed values and predicted values for the nonsampled
units.  In this situation, it may be more convenient to use a completely model-based
approach for prediction of the small area means \citep{bat88}.


\section{Unweighted analysis }\label{sec: uw}

\subsection{Ignorable design}

First, assume the survey design is {\em ignorable} or {\it noninformative}.
Ignorable designs, such as simple random sampling with replacement, arise when the
sample inclusion variable \(I\) is independent of the response variable \(y\).  In
this situation, the distribution of the sampled responses will be identical to the
distribution of nonsampled responses.  That is, if a model \(f( \cdot \mid
\boldsymbol{\theta} )\) is assumed to hold for all nonsampled units in the
population, then it will also hold for the sampled units, since the {\em sample
  distribution} of \(y\), \(f(y \mid I = 1, \boldsymbol{\theta}) = f( y \mid
\boldsymbol{\theta})\) is identical to the population distribution of \(y\).  In this
case, a model can be fit to the sampled data, and the fitted model can then be used
directly to predict the nonsampled units, without needing any adjustments due to the
survey design.


The nested error regression model or, using the terminology of \citet{rao15}, the
basic unit-level model, was introduced by \citet{bat88} for estimation of small area
means using data obtained from a survey with an ignorable design.  Consider the
linear mixed effects model
\begin{equation}\label{E: bhf}
  y_{ij} = \boldsymbol{x}_{ij} \boldsymbol{\beta} + v_i + e_{ij},
\end{equation}
where \(i = 1, \dots, m\) indexes the different small areas of interest, and \(j \in
\mathcal{S}_i\) indexes the sampled units in small area \(i\).  Here, the model
errors, \(v_i\), are i.i.d. \(N \left( 0, \sigma^2_v \right)\) random variables, and
the sampling errors, \(e_{ij}\), are i.i.d. \(N \left( 0, \sigma^2_e \right)\) random
variables, independent of the model errors.

Let \(\boldsymbol{V}_i\) be the covariance matrix consisting of diagonal elements
\(\sigma^2_v + \sigma^2_e/n_i\), and off-diagonal elements \(\sigma^2_v\).  Assuming
\eqref{E: bhf} holds for the sampled units, and the variance parameters
\(\sigma^2_v\) and \(\sigma^2_e\) are known, the best linear unbiased predictor
(BLUP) of \(\bar{y}_i = \sum_{j \in \mathcal{U}_i} y_{ij} / N_i\) is
\begin{equation}\label{E: bhfBLUP}
  \hat{\bar{y}}_i = \frac{1}{N_i} \sum_{j \in \mathcal{S}_i} y_{ij} + \frac{1}{N_i}
    \sum_{j \in \mathcal{S}^c_i} \left( \boldsymbol{x}^T_{ij}
    \tilde{\boldsymbol{\beta}} + \tilde{v}_i \right),
\end{equation}
where
\begin{equation*}
  \tilde{\boldsymbol{\beta}} = \left( \sum^m_{i = 1} \boldsymbol{X}^T_i
    \boldsymbol{V}^{-1}_i \boldsymbol{X}_i \right)^{-1} \left( \sum^m_{i = 1}
    \boldsymbol{X}^T_i \boldsymbol{V}^{-1}_i \boldsymbol{y}_i \right),
\end{equation*}
\(\boldsymbol{X}_i\) is the \(n_i \times p\) matrix with rows
\(\boldsymbol{x}^T_{ij}\), and \(\tilde{v}_i = (\sigma^2_v / (n_i \sigma^2_v +
\sigma^2_e )) \sum_{j \in \mathcal{S}_i} (y_{ij} - \boldsymbol{x}^T_{ij}
\tilde{\boldsymbol{\beta}} )\).  In \eqref{E: bhfBLUP}, as in the general expression
in \eqref{E: bp}, the unobserved \(y_{ij}\) are replaced by model predictions.  Note
that evaluation of \eqref{E: bhfBLUP} requires knowledge of the population mean,
\(\bar{\boldsymbol{X}}_{ip} = \sum_{j \in \mathcal{U}_i} \boldsymbol{X}_{ij} / N_i\),
of the covariates.


In practice, the variance components \(\sigma^2_v\) and \(\sigma^2_e\) are unknown
and need to be estimated.  The empirical best linear unbiased predictor (EBLUP) is
obtained by substituting estimates, \(\hat{\sigma}^2_e\) and \(\hat{\sigma}^2_v\),
(typically MLE, REML, or moment estimates) of the variance components in the above
expressions \citep{pra90}. In addition, this model could easily be fit using Bayesian
hierarchical modeling rather than using the EBLUP, which would incorporate the
uncertainty from the variance parameters.  \citet{mol14} developed a Bayesian version
of the nested error regression model (\ref{E: bhf}), using noninformative priors on the
variance components.

The survey weights do not enter into either the nested error regression model
\eqref{E: bhf} or the EBLUPs of the small area means \eqref{E: bhfBLUP}.  Because of
this, the EBLUP is not design-consistent, unless the sampling design is
self-weighting within each small area \citep{rao15}.

\subsection{Including design variables in the model}

Suppose now that the survey design is informative, so that the way in which
individuals are selected in the sample depends in an important way on the value of
the response variable \(y_{ij}\).  It is well established that when the survey design
is informative, that ignoring the survey design and performing unweighted analyses
without adjustment can result in substantial biases \citep{nat80, pfe07}.

One method to eliminate the effects of an informative design is to condition on all
design variables \citep[][Chap. 7]{gel95}.  To see this, decompose the
response variables as \(\boldsymbol{y} = ( \boldsymbol{y}_s, \boldsymbol{y}_{ns})\),
where \(\boldsymbol{y}_s\) are the observed responses for the sampled units in the
population, and \(\boldsymbol{y}_{ns}\) represents the unobserved variables
corresponding to nonsampled individuals.  Let \(\boldsymbol{I}\) be the matrix of
sample inclusion variables, so that \(I_{ij} = 1\) if \(y_{ij}\) is observed and
\(I_{ij} = 0\) otherwise.  The observed data likelihood, conditional on covariate
information \(\boldsymbol{X}\), and model parameters \(\boldsymbol{\theta}\) and
\(\boldsymbol{\phi}\), is then
\begin{equation*}
  f( \boldsymbol{y}_s, \boldsymbol{I} \mid \boldsymbol{X}, \boldsymbol{\theta},
    \boldsymbol{\phi} ) = \int f(\boldsymbol{y}_s, \boldsymbol{y}_{ns},
    \boldsymbol{I} \mid \boldsymbol{X}, \boldsymbol{\theta}, \boldsymbol{\phi} )
    d \boldsymbol{y}_{ns} \\
  = \int f ( \boldsymbol{I} \mid \boldsymbol{y}, \boldsymbol{X},
    \boldsymbol{\phi} ) f ( \boldsymbol{y} \mid \boldsymbol{X},
    \boldsymbol{\theta} ) d \boldsymbol{y}_{ns}.
\end{equation*}
If \(f ( \boldsymbol{I} \mid \boldsymbol{y}, \boldsymbol{X}, \boldsymbol{\phi} ) = f
( \boldsymbol{I} \mid \boldsymbol{X}, \boldsymbol{\phi} )\), the inclusion variables
\(\boldsymbol{I}\) are independent of \(\boldsymbol{y}\), conditional on
\(\boldsymbol{X}\), and the survey design can be ignored.  For example, if the design
variables \(\boldsymbol{Z}\) are included in \(\boldsymbol{X}\), the ignorability
condition may hold, and inference can be based on \(f ( \boldsymbol{y}_s \mid
\boldsymbol{X}, \boldsymbol{\theta})\).  


\citet{little12} advocates for a general framework using unit-level Bayesian modeling
that incorporates the design variables.  For example, if cluster sampling is used,
one could incorporate a cluster level random effect into the model, or if a
stratified design is used, one might incorporate fixed effects for the strata.  The
idea is that when all design variables are accounted for in the model, the
conditional distribution of the response given the covariates for the sampled units
is independent of the inclusion probabilities.  Because the model is unit-level and
Bayesian, the unsampled population can be generated via the posterior predictive
distribution.  Doing so provides a distribution for any finite population quantity
and incorporates the uncertainty in the parameters. For example, if the population
response is generated at draw $k$ of a Markov chain Monte Carlo algorithm,
$\bm{y}^{{(k)}}$, then one has implicitly generated a draw from the
posterior distribution of the population mean for a given area $i$:
$$ \bar{y}^{{(k)}}_i = \frac{\sum_{j=1}^N y_j^{{(k)}} I\left(
  j \in \mathcal{U}_i \right) }{ \sum_{j=1}^N I \left(j \in \mathcal{U}_i
  \right)}. $$
If there are $K$ total posterior draws, one could then estimate the mean and standard
error of $\bar{y}_i$ with
$$\hat{\bar{y}}_i = \frac{1}{K} \sum_{j=1}^K
  \bar{y}^{{(k)}}_i$$
and
$$\widehat{SE(\hat{\bar{y}}_i)} =
  \sqrt{Var(\hat{\bar{y}}_i)}.$$

The problem with attempting to eliminate the
effect of the design by conditioning on design variables is often more of a practical
one, because neither the full set of design variables, nor the functional
relationship between the design and the response variables will be fully known.
Furthermore, expanding the model by including sufficient design information so as to
ignore the design may make the likelihood extremely complicated or even intractable.

\subsection{Poststratification}
  
\citet{little93} gives an overview of poststratification. To perform
poststratification, the population is assumed to contain \(m\) categories, or
poststratification cells, such that within each category units are independent and
identically distributed.  Usually these categories are cross-classifications of
categorical predictor variables such as county, race, and education level. When a
regression model is fit relating the response to the predictors, predictions can be
generated for each unit within a cell, and thus for the entire
population. Importantly, any desired aggregate estimates can easily be generated from
the unit level population predictions.

\citet{gelman97} and \citet{park06} develop a framework for poststratification via
hierarchical modeling. By using a hierarchical model with partial pooling, parameter
estimates can be made for poststratification cells without any sampled units, and
variance is reduced for cells having few sampled units. \citet{gelman97} and
\citet{park06} provide an example for binary data that uses the following model
\begin{equation}\label{E: ps}
  \begin{split}
    y_{ij}\vert p_{ij} & \sim Bernoulli(p_{ij}) \\
    \mbox{logit}(p_{ij}) &= \boldsymbol{x}_{ij}' \boldsymbol{\beta} \\
    \boldsymbol{\beta} &= (\boldsymbol{\gamma}_1, \ldots , \boldsymbol{\gamma}_K) \\
    \boldsymbol{\gamma}_k & \stackrel{ind}{\sim} N_{c_k}(0, \sigma^2_k
      \boldsymbol{I}_{c_k}),k=1,\ldots,K,
  \end{split}
\end{equation}
where \(\boldsymbol{x}_{ij}\) is a vector of dummy variables for $K$ categorical
predictor variables with \(c_k\) classes in variable $k$.
  
Bayesian inference can be performed on this model, leading to a probability,
\(p_{ij}=p_{i}, \forall j\), that is constant within each cell \(i=1,\ldots,m\).  The
number of positive responses within cell \(i\) can be estimated with \(N_i p_i\), and
any higher level aggregate estimates can be made by aggregating the corresponding
cells.  In some scenarios, the number of units within each cell may not be known, in
which case further modeling would be necessary.

\section{Models with survey weight adjustments}\label{sec: pseudo}

Although many of the the models in Section \ref{sec: uw} can be used to handle informative sampling, they do not rely on the survey weights. In this section, we explore techniques that rely on the weights to adjust the sample likelihood.

There have been several methods proposed in the literature which make use of the nested error
regression model \eqref{E: bhf}, but which incorporate the survey weights, either
as regression variables, or as adjustments to the predicted values, so as to
protect against a possible informative survey design.

\subsection{Survey weight adjustments to the basic unit-level model}

\citet{ver15} augmented the nested error regression model \eqref{E: bhf}, by including functions of the inclusion probabilities, \(g \left( \pi_{ij} \right)\), as predictors.  Care must be taken in the choice of the function \(g\), as knowledge of the population means \(\bar{G}_i = \sum_{i \in \mathcal{U}_j} g \left( \pi_{ij} \right) / N_i\) need to be known to obtain the EBLUPs from
\eqref{E: bhfBLUP}.  Some suggestions for the choice of \(g\) were \(g \left( \pi_{ij} \right) = \pi_{ij}\), which gives \(\bar{G}_i = 1/N_i\), and \(g \left( \pi_{ij} \right) = n_i / \pi_{ij}\), which gives \(\bar{G}_i = n_i \sum_{j \in \mathcal{U}_i} w_{ij} / N_i\), which may be known in practice.  \citet{ver15} reported strong performance of the EBLUP using the augmented nested error
regression model, in a probability proportional to size simulation study, in terms of bias and mean squared error, for properly chosen augmenting variable \(g\).  However, some choices of \(g\), such as \(g \left( \pi_{ij} \right) = w_{ij}\), could lead to poor performance, except under non-informative sampling.  \citet{ver15} suggested using scatter plots of residuals from the nested error
regression model against different choices of augmenting variables to choose an appropriate model.  An alternative to exploring a collection of augmenting variables is to estimate the functional form of \(g\).  \citet{zhe03} investigated nonparametric estimation of \(g\) using penalized splines, and found that predictions of small area means using this modeling framework resulted in large gains in mean squared error over the design-based estimates in their simulation studies.

\citet{you02} proposed a pseudo-EBLUP of the small area means \(\theta_i =
\bar{\boldsymbol{X}}^T_i \boldsymbol{\beta} + \nu_i\) based on the nested error
regression model \eqref{E: bhf}, which incorporates the survey weights.  In their
approach, the regression parameters \(\boldsymbol{\beta}\) in \eqref{E: bhf} are
estimated by solving a system of survey-weighted estimating equations
\begin{equation}\label{E: psEE}
  \sum^m_{i = 1} \sum_{j \in \mathcal{S}_i} w_{ij} \boldsymbol{x}_{ij} \{ y_{ij} -
    \boldsymbol{x}^T_{ij} \boldsymbol{\beta} - \gamma_{iw} ( \bar{y}_{iw} -
    \bar{\boldsymbol{x}}^T_{iw} \boldsymbol{\beta} ) \} = 0,
\end{equation}
where \(\gamma_{iw} = \sigma^2_\nu/\left( \sigma^2_\nu + \sigma^2_\epsilon \delta^2_i
\right)\), \(\delta^2_i = \sum_{j \in \mathcal{S}_i} w^2_{ij}\), \(\bar{y}_{iw} =
\sum_{j \in \mathcal{S}_i} w_{ij} y_{ij} / \sum_{j \in \mathcal{S}_i} w_{ij}\), and
\(\bar{\boldsymbol{x}}_{ij} = \sum_{j \in \mathcal{S}_i} w_{ij} \boldsymbol{x}_{ij} /
\sum_{j \in \mathcal{S}_i} w_{ij}\).  This is an example of the pseudo-likelihood
approach to incorporating survey weights, which is later discussed in more detail.

The pseudo-BLUP \(\tilde{\boldsymbol{\beta}}_w = \tilde{\boldsymbol{\beta}}_w (
\sigma^2_e, \sigma^2_v)\) is the solution to \eqref{E: psEE} when the variance
components \(\sigma^2_e\) and \(\sigma^2_v\) are known, and the pseudo-EBLUP,
\(\hat{\boldsymbol{\beta}}_w = \tilde{\boldsymbol{\beta}} (\hat{\sigma}^2_e,
\hat{\sigma}^2_v)\), is the solution to \eqref{E: psEE} using plug-in estimates
\(\hat{\sigma}^2_e\) and \(\hat{\sigma}^2_v\) of the variance components.  The
pseudo-EBLUP, \(\hat{\theta}_i\) of the small area mean \(\theta_i\) is then
\begin{equation*}
  \hat{\theta}_{iw} = \hat{\gamma}_{iw} \bar{y}_{iw} + \left( \bar{X}_i -
    \hat{\gamma}_{iw} \bar{x}_{iw} \right)^T \hat{\boldsymbol{\beta}}_w.
\end{equation*}

Similar to \citet{bat88}, \citet{you02} assumed an ignorable survey design, so that
the model \eqref{E: bhf} holds for both the sampled and nonsampled units.  However,
\citet{you02} showed that inclusion of the survey weights in the pseudo-EBLUP results
in a design-consistent estimator.  In addition, when the survey weights are
calibrated to the population total, so that \(\sum_{j \in \mathcal{S}_i} w_{ij} =
N_i\), the pseudo-EBLUP has a natural benchmarking property, without any additional
adjustment, in the sense that
\begin{equation*}
  \sum^m_{i = 1} N_i \hat{\theta}_{iw} = \hat{Y}_w + \left( \boldsymbol{X} -
    \hat{\boldsymbol{X}}_w \right)^T \hat{\boldsymbol{\beta}}_w,
\end{equation*}
where \(\hat{Y}_w = \sum^m_{i = 1} \sum_{j \in \mathcal{S}_i} w_{ij} y_{ij}\) and
\(\hat{\boldsymbol{X}}_w = \sum^m_{i = 1} \sum_{j \in \mathcal{S}_i} w_{ij}
\boldsymbol{x}_{ij}\).  That is, the weighted sum of area-level pseudo-EBLUPs is
equal to a GREG estimator of the population total.

 An alternative pseudo-EBLUP, which is applicable to estimation of
  general small area parameters beyond the small area means, was proposed in
  \citet{gua18} \citep[see also][]{jia06}.  Rather than use the genuine best
  predictor in \eqref{E: bp}, which conditions on all observed data, \citet{gua18}
  suggested a pseudo-best predictor, which conditions only on the survey-weighted
  Horvitz-Thompson estimator, \(\bar{y}_{iw} = \sum_{j \in \mathcal{S}_i} w_{ij}
  y_{ij} / \sum_{j \in \mathcal{S}_i} w_{ij}\), of the small area means.  Assuming
  that the nested error regression model \eqref{E: bhf} holds for all units in the
  population, there is a simple, closed-form expression for the predictions of
  out-of-sample variables, \(y_{ij}\), given by
  \begin{equation*}
    E \left( y_{ij} \mid \bar{y}_{iw} \right) = \boldsymbol{x}^T_{ij}
      \boldsymbol{\beta} + \gamma_{iw} \left( \bar{y}_{iw} -
      \bar{\boldsymbol{x}}^T_{iw} \boldsymbol{\beta} \right),
  \end{equation*}
  using the same notation as in \eqref{E: psEE}.  This idea can easily be extended
  for prediction of general additive parameters, \(H_i = \sum_{j \in \mathcal{U}_i}
  h(y_{ij}) / N_i\), by using the conditional expectation \(E \left( h(y_{ij}) \mid
  \bar{y}_{iw} \right)\) in place of the out-of-sample variables.

\subsection{Pseudo-likelihood approaches}

Suppose the finite population values \(y_i\), are independent, identically
distributed realizations from a known superpopulation distribution, \(f_p (y \mid
\boldsymbol{\theta})\).  Here, for notational convenience, we use a single subscript
\(i\) to index the finite population.
Standard likelihood analysis for inference on \(\boldsymbol{\theta}\), using only the
observed sampled values, could produce asymptotically biased estimates when the
sampling design is informative \citep{pfe98b}.  Pseudo-likelihood analysis,
introduced by \citet{bin83} and \citet{ski89}, incorporates the survey weights into
the likelihood for design-consistent estimation of \(\boldsymbol{\theta}\).

The pseudo-log-likelihood is defined as
\begin{equation}\label{E: pseudoLogLikelihood}
  \sum_{i \in \mathcal{S}} w_i \log f_p( y_i \mid \boldsymbol{\theta});
\end{equation}
this is simply the Horvitz-Thompson estimator of the population-level log likelihood.
Inference on \(\boldsymbol{\theta}\) can be based on the maximizer,
\(\hat{\boldsymbol{\theta}}_{PS}\) (designating the maximum of the pseudo-likelihood
rather than the likelihood), of \eqref{E: pseudoLogLikelihood}, or equivalently, by
solving the system
\begin{equation}\label{E: pseudoDLogLikelihood}
  \sum_{i \in \mathcal{S}} w_i \frac{\partial}{\partial \boldsymbol{\theta}} \log f_p
    (y_i \mid \boldsymbol{\theta}) = \boldsymbol{0}.
\end{equation}

The system \eqref{E: pseudoDLogLikelihood} is an example of the use of
survey-weighted estimating functions for inference on a superpopulation parameter
\citep{bin83, bin94}.  More generally, let \(\boldsymbol{\theta}_N =
\boldsymbol{\theta}_N \left( \left\{ y_i \right\} \right)\) be a superpopulation
parameter of interest, which is a function of the finite population values \(y_i, i =
1, \dots, N\), that can be obtained as a solution to a ``census'' estimating equation
\begin{equation}\label{E: EE}
  \boldsymbol{\Phi} \left( \mathbf{y}; \boldsymbol{\theta} \right) = \sum^N_{i = 1}
    \boldsymbol{\phi}_i ( y_i; \boldsymbol{\theta} ) = \boldsymbol{0},
\end{equation}
where the \(\boldsymbol{\phi}_i\) are known functions of the data and the parameter,
with mean zero under the superpopulation model.  The term ``census'' is used to
describe the estimating function \eqref{E: EE}, because \eqref{E: EE} can only be
calculated if all finite population values are observed, or if a census of the
population is conducted.

The target parameter \(\boldsymbol{\theta}_N\) is defined implicitly as a solution to
the census estimating equation \eqref{E: EE}.  A point estimate,
\(\hat{\boldsymbol{\theta}}_N\), of \(\boldsymbol{\theta}_N\) can be obtained by
finding a root of a design-unbiased estimate, \(\hat{\boldsymbol{\Phi}} \left(
\boldsymbol{y}_s; \boldsymbol{\theta} \right)\), of \(\boldsymbol{\Phi}\), such as
the Horvitz-Thompson estimator
\begin{equation}\label{E: wtEE}
  \hat{\boldsymbol{\Phi}} \left( \boldsymbol{y}_s, \boldsymbol{\theta} \right) =
    \sum_{i \in \mathcal{S}} w_i \boldsymbol{\phi}_i \left( y_i; \boldsymbol{\theta}
    \right) = \boldsymbol{0}.
\end{equation}


The use of an estimating function, rather than the score function, can be
advantageous, as it reduces the number of assumptions about the superpopulation that
need to be made.  Full distributional specification is not required, and instead only
assumptions about the moment structure are needed.  The choice of the specific
estimating function may be motivated by a conceptual superpopulation model, or a
finite population parameter of interest.  Regardless of whether a superpopulation
model is assumed, most finite population parameters of interest can be formulated as
a solution to a census estimating equation, and well-known `model assisted'
estimators of the finite population parameters can be derived as solutions of
survey-weighted estimating equations.  For example, the estimating function \(\phi
(y_i; \theta) = (y_i - \theta)\) leads to the population total, \(\sum_{i \in
  \mathcal{U}} y_i\), and its estimator \(\sum_{i \in \mathcal{S}} w_i y_i/ \sum_{i
  \in \mathcal{S}} w_i\).  If \(t_x = \sum_{i \in \mathcal{U}} x_i\) is known, the
pair of estimating functions \(\phi_1 (y_i; x_i, \theta_1) = (y_i - x_i \theta_1)\)
and \(\phi_2 (y_i, \theta_2) = (y_i - t_x \theta_2)\), give the population total
\(t_y = \sum_{i \in \mathcal{U}} y_i\) and its ratio estimator \(t_x \sum_{i \in
  \mathcal{S}} w_i y_i / \sum_{i \in \mathcal{S}} w_i x_i\).  If the covariate vector
\(\boldsymbol{x}_i\) contains an intercept, the pair of estimating functions \(\phi_1
( y_i; \theta_1, \boldsymbol{\theta}_2 ) = (\theta_1 - t_x \boldsymbol{\theta}_2)\)
and \(\boldsymbol{\phi}_2 = (y_i - \boldsymbol{x}^T_i \boldsymbol{\theta}_2)
\boldsymbol{x}_i\) leads to the finite population total and its GREG estimator
\(\sum_{i \in \mathcal{U}} \boldsymbol{x}^T_i \boldsymbol{\theta}_2\), where
\(\boldsymbol{\theta}_2 = (\sum_{i \in \mathcal{S}} w_i \boldsymbol{x}_i
\boldsymbol{x}^T_i)^{-1} \sum_{i \in \mathcal{S}} w_i y_i \boldsymbol{x}_i\).

An important aspect of small area modeling is the introduction of area specific
random effects to link the different small areas and to ``borrow strength,'' by
relating the different areas through the linking model, and introducing auxiliary
covariate information, such as administrative records.  The presence of random
effects and the multilevel structure of small area models means that neither the
pseudo-likelihood method nor the related estimating function approach can be directly
applied to small area estimation problems.  However, \citet{gri04}, \citet{asp06},
\citet{rab06} extended the pseudo-likelihood approach to  accommodate models with
hierarchical structure.

Let \(v_i \overset{i.i.d.}{\sim} \varphi (v)\) denote the area specific random
effects with common density \(\varphi\).  The usual choice for \(\varphi\) is the
mean zero normal distribution with unknown variance \(\sigma^2\).  Suppose now that
the finite population \(y_{i1}, \dots, y_{i N_i}\) in small areas \(i = 1, \dots, m\)
are i.i.d. realizations from the superpopulation \(f_i (y \mid \boldsymbol{\theta},
v_i)\), and let \(\mathcal{S}_i\) be the sampled units in area \(i\).  The census
marginal log-likelihood is obtained by integrating out the random effects from the
likelihood:
\begin{equation}\label{E: multLogLik}
  \begin{split}
    \log L (\boldsymbol{\theta}) & = \sum^m_{i = 1} \log \int \prod_{j \in
      \mathcal{U}_i} f_i (y_{ij} \mid \boldsymbol{\theta}, v) \varphi (v) dv \\
    \hspace{20mm} &= \sum^m_{i = 1} \log \int \exp \left\{ \sum_{j \in \mathcal{U}_i}
      \log f_i (y_{ij} \mid \boldsymbol{\theta}, v) \right\} \varphi (v) dv.
  \end{split}
\end{equation}

Suppose the survey weights are decomposed into two components, \(w_{j \mid i}\), and
\(w_i\), where \(w_{j \mid i}\) is the weight for unit \(j\) in area \(i\), given
that area \(i\) has been sampled, and \(w_i\) is the weight associated to small area
\(i\).  The pseudo-log-likelihood for the multilevel model can be defined by
replacing \(\sum_{j \in \mathcal{U}_i} \log f_i (y_{ij} \mid \boldsymbol{\theta},
v_i)\) in \eqref{E: multLogLik} by the design-unbiased estimate, \(\sum_{j \in
  \mathcal{S}_i} w_{j \mid i} \log f_i (y_{ij} \mid \boldsymbol{\theta}, v_i)\), to
get
\begin{equation}\label{E: multPseudoLogLikelihood}
  \log \hat{L} (\boldsymbol{\theta}) = \sum^m_{i = 1} w_i \log \int \exp \left\{
    \sum_{j \in \mathcal{S}_i} w_{j \mid i} \log f(y_{ij} \mid \boldsymbol{\theta},
    v) \right\} \varphi (v) dv.
\end{equation}

 Analytical expressions for the maximizer of \eqref{E:
    multPseudoLogLikelihood} generally do not exist, so the maximum pseudo-likelihood
  estimator, \(\hat{\boldsymbol{\theta}}_{ps}\), must be found by numerical
  maximization of \eqref{E: multPseudoLogLikelihood}.  \citet{gri04} used the {\tt
    NLMIXED} procedure within SAS, using appropriately adjusted weights in the {\tt
    replicate} statement and a bootstrap for mean squared error estimation.
  \citet{rab06} used an adaptive quadrature routine using the {\tt gllamm} program
  within Stata, and derived a sandwich estimator of the standard errors, finding good
  coverage in their simulation studies with this estimate.
  \citet{kim2017statistical} proposed an EM algorithm for parameter estimation.
  Their method involves two steps, where first the random effects are treated as
  fixed, and a profile likelihood maximum likelihood estimator of the random effects
  are computed.  The second step uses the EM algorithm to estimate the remaining
  model parameters.  Their method relies on a normal approximation to the predictive
  distribution of the random effects, but was found to give good results with
  moderate cluster sizes in numerical studies.  \citet{kim2017statistical} also gave
  a method for predicting random effects, using the EM algorithm and an approximating
  predictive distribution that was shown to be valid for sufficiently large cluster
  sizes, which is needed for prediction of unobserved variables. 

\citet{rao13} noted that both design consistency and design-model consistency of
\(\hat{\boldsymbol{\theta}}_{ps}\) as an estimator of the finite population parameter
\(\boldsymbol{\theta}_N\), or the model parameter \(\boldsymbol{\theta}\),
respectively, requires that both the number of areas (or clusters), \(m\), and the
number of elements within each cluster, \(n_i\), tend to infinity, and that the
relative bias of the estimators can be large when the \(n_i\) are small.
\cite{rao13} showed that consistency can be achieved with only \(m\) tending to
infinity (allowing the \(n_i\) to be small) if the joint inclusion probabilities,
\(\pi_{jk \mid i}\), are available.  Their method is to use the marginal joint
densities \(f(y_{ij}, y_{ik} \mid \boldsymbol{\theta})\), of elements in a cluster,
integrating out the random effects, in the pseudo-log likelihood, and to estimate
\(\boldsymbol{\theta}\) by maximizing the design-weighted pseudo log likelihood
\begin{equation}\label{E: wComp}
  l_{wC} (\boldsymbol{\theta}) = \sum_{i \in \mathcal{S}} w_i \sum_{j < k \in
    \mathcal{S}_i} w_{jk \mid i} \log f(y_{ij}, y_{ik} \mid \boldsymbol{\theta}).
\end{equation}
It was shown in \citet{yi16} that the maximizer of \eqref{E: wComp},
\(\boldsymbol{\theta}_{wC}\), is consistent for the second-level parameters
\(\boldsymbol{\theta}\), with respect to the joint superpopulation model and the
sampling design.

There are two important considerations when using the pseudo-likelihood in multilevel
models.  The first is that two sets of survey weights, \(w_i\) and \(w_{j\mid i}\)
(and in the case of the method of \citet{rao13}, higher-order inclusion
probabilities, \(\pi_{jk\mid i}\)) are required, which is not typically the case;
access to only the joint survey weights \(w_{ij}\) is not sufficient to use the
multilevel models, unless all clusters \(i = 1, \dots, m\) sampled with certainty.
The second consideration is that use of unadjusted, second level weights \(w_{j\mid
  i}\) can cause significant bias in estimates of variance components. For
single level models, scaling the weights by any constant factor does not change
inference, as the solution to \eqref{E: pseudoDLogLikelihood} is clearly invariant to
any scaling of the weights.  However, for multilevel models, the maximum
pseudo-likelihood estimator and the associated mean squared prediction error may
change depending on how the weights \(w_{j \mid i}\) are scaled.  Some suggestions
include using scaled weights \(\tilde{w}_{j \mid i}\) such that: 1) \(\sum^{n_i}_{j =
  1} \tilde{w}_{j\mid i} = n_i\) \citep{asp06, gri04, pfe98b}, 2) constant scaling
across clusters, so that \(\sum^{n_i}_{j = 1} \tilde{w}_{j\mid i} = \sum^{n_i}_{j =
  1} w_{j \mid i}\) \citep{asp06}, 3) \(\sum^{n_i}_{j = 1} \tilde{w}_{i\mid j} =
n^*_i\), where \(n^*_i = (\sum_j w_{i\mid j})^2 / \sum_j w^2_{ij}\) is the effective
sample size in cluster \(i\) \citep{pot92, pfe98b, asp06}, and 4) unscaled
\citep{pfe98b, gri04, asp06}. However, \citet{kor03} showed that any scaling method can produce seriously
biased variance estimates under informative sampling schemes, even with large sample
sizes in each cluster.  There does not seem to be a single `best' scaling method that can be used without consideration of the sampling scheme,
or the working likelihood.

The pseudo-log-likelihoods \eqref{E: pseudoLogLikelihood} and \eqref{E:
  multPseudoLogLikelihood} suggest pseudo-likelihoods
\begin{equation}\label{E: pseudoLikelihood}
  \prod^m_i \prod_{j \in \mathcal{S}_i} f(y_{ij} \mid \boldsymbol{\theta})^{w_{ij}}
\end{equation}
for single level models, and
\begin{equation}\label{E: multPseudoLikelihood}
  \prod^m_{i = 1} \left\{ \int \prod_{j \in \mathcal{S}_i} f (y_{ij} \mid
    \boldsymbol{x}_{ij}, \boldsymbol{\theta}, v_j)^{w_{j\mid i}} \phi ( v_j ) d v_j
    \right\}^{w_i}
\end{equation}
for multilevel models \citep{asp06}.  The pseudo-likelihood \eqref{E:
  pseudoLikelihood} is sometimes called the composite likelihood in general
statistical problems, when the weights \(w_{ij}\) (not necessarily survey weights)
are known positive constants, and its use is popular in problems where the exact
likelihood is intractable or computationally prohibitive \citep{var11}.

The pseudo-likelihood \eqref{E: pseudoLikelihood} is not a genuine likelihood, as it
does not incorporate the dependence structure in the sampled data nor the
relationship between the responses and the design variables beyond inclusion of the
survey weights.  However, the pseudo-likelihood has been shown to be a useful tool
for likelihood analysis for finite population inference in both the frequentist and
Bayesian framework.

By treating the pseudo-likelihood as a genuine likelihood, and specifying a prior
distribution \(\pi (\boldsymbol{\theta})\) on the model parameters
\(\boldsymbol{\theta}\), Bayesian inference can be performed on
\(\boldsymbol{\theta}\).  For general models, \citet{sav16} showed for certain
sampling schemes, and for a class of population distributions, that the
pseudo-posterior distribution using the survey weighted pseudo-likelihood, with
survey weights scaled to sum to the sample size, \eqref{E: pseudoLikelihood}
converges in \(L^1\) to the population posterior distribution.  This result justifies
use of \eqref{E: pseudoLikelihood} in place of the likelihood in Bayesian analysis of
population parameters, conditional on the observed sampled units, even when the
sample design is informative.  Predictions of area-level random effects as well of
predictions of nonsampled units can then be made as well.

The authors focus on parameter inference and do not give any advice for making
area-level estimates.  However, it is straightforward to implement a model with a Bayesian
pseudo-likelihood and then apply poststratification after the fact by generating the
population, and thus any desired area-level estimates using \eqref{E: bp}. This type
of pseudo-likelihood with poststratification for SAE was demonstrated in the
frequentist setting by \citet{zha14}.

\citet{rib12} provides a discussion on the validity of Bayesian inference using the
composite likelihood \eqref{E: pseudoLikelihood} in place of the exact likelihood in
Bayes' formula.  An example of this method used in the sample survey context can be
found in \citet{don14}, which used a weighted pseudo-likelihood with a multinomial
distribution as a model for binned response variables.  They assumed an improper
Dirichlet distribution on the cell probabilities, and used the associated posterior
and posterior predictive distributions for prediction of the nonsampled population
units.

In a similar spirit, \citep{rao10} use a Bayesian pseudo-empirical
  likelihood to create estimates for a population mean. They form the
  pseudo-empirical likelihood function as
\begin{equation*} 
  L_{PEL}(p_1, \ldots , p_n) = \prod_{i \in \mathcal{S}} p_i^{\tilde{w_i}}
\end{equation*}
where the weights are scaled to sum to the sample size. They accompany this with a
Dirichlet prior distribution over \( (p_1, \ldots, p_n) \), and thus conjugacy yields
a Dirichlet posterior distribution
\begin{equation*}
  \pi(p_1, \ldots, p_n | D_S) = c(\tilde{w_1} + \alpha_1, \ldots , \tilde{w_n} +
    \alpha_n) \prod_{i \in \mathcal{S}} p_i^{\tilde{w_i}+ \alpha_i -1},
\end{equation*}
where \(c\) represents the normalizing constant.  The posterior distribution of the
population mean, \( \theta \), corresponds to the posterior distribution of \(\sum_{i
  \in \mathcal{S}} p_i y_i \). It is straightforward to use Monte Carlo techniques to
sample from this posterior.  The authors also note that the design weights can be
replaced with calibration weights in order to include auxiliary variables.

\subsection{Regressing on the Survey Weights}

Prediction of small area quantities using \eqref{E: bp} requires estimation of
\(E(y_{ij} \mid D_s)\) for all nonsampled units in the population.  One of the main
difficulties in using unit-level model-based methods is the lack of knowledge of the
covariates, sampling weights, or population sizes associated with the nonsampled
units and small areas, that are needed to make these model-based predictions.  To
overcome this difficulty, \citet{si15} modeled the observed poststratification cells
\(n_i\), conditional on \(n = \sum^m_{i = 1} n_i\), using the multinomial
distribution
\begin{equation}
    (n_1,\ldots,n_m) \sim \hbox{Multinomial} \left(n ; \frac{N_1/w_1}{\sum_{i=1}^m
      N_i/w_i },\ldots,\frac{N_m/w_m}{\sum_{i=1}^m N_i/w_i }\right)
\end{equation}
for poststratification cells \(i = 1, \ldots, m\).

This model assumes that the unique values of the sample weights determine the
poststratification cells, and that the sampling weight and response are the only
values known for sampled units. The authors state that, in general, this assumption is
untrue, because there will be cells with low probability of selection that do not
show up in the sample, but the assumption is necessary in order to proceed with the
model.  This model yields a posterior distribution over the cell population sizes
which can be used for poststratification with their response model, which uses a
nonparametric Gaussian process regression on the survey weights,
\begin{equation}
  \begin{split}
    y_{ij} | \mu(w_i), \sigma^2 &\sim \hbox{N}(\mu(w_i), \sigma^2) \\ \mu(w_i) | \beta,
      C(w_i,w_{i'}|\boldsymbol{\theta}) &\sim \hbox{GP}(w_i\beta, C(w_i,
      w_{i'}|\boldsymbol{\theta})) \\ &\pi(\sigma^2, \beta, \boldsymbol{\theta}),
  \end{split}
\end{equation}
for observation \(j\) in poststratification cell \(i\). Here,
\(C(\cdot,\cdot|\boldsymbol{\theta})\) represents a valid covariance function that
depends on parameters \(\boldsymbol{\theta}\). The authors use a squared exponential
function, but other covariance functions could be used in its place. The normal
distribution placed over \(y_{ij}\) could be replaced with another distribution in
the case of non-Gaussian data. Specifically, the authors explore the Bernoulli
response case. This model implicitly assumes that units with similar weights will
tend to have similar response values, which is likely not true in general. However,
in the absence of any other information about the sampled units, this may be the most
practical assumption. Because \citet{si15} assume that only the survey weights and
response values are known, this methodology cannot be used for small area estimation
as presented. However, the model can be extended to include other variables such as
county, which would allow for area level estimation.

\citet{van16} extend the work of \citet{si15} to be applied to small area
estimation. They assume that the poststratification cells are designated by the
unique weights within each area. Rather than using the raw weights, they use the
weights scaled to sum to the sample size within each area. They then use a similar
multinomial model to \citet{si15} in order to perform poststratification using the
posterior distribution of the poststratification cell population sizes. Assuming a
Bernoulli response, they use the data model
\begin{equation}
  \begin{split}
    y_{ij} | \eta_{ij} &\sim \hbox{Bernoulli}(\eta_{ij}) \\
    \hbox{logit}(\eta_{ij}) &= \beta_0 + \mu(\stackrel{\sim}{w}_{ij}) + u_i + v_i
  \end{split}
\end{equation}
for unit \(j\) in small area \(i\), with \(\stackrel{\sim}{w}_{ij}\) designating the
scaled weights. Independent area level random effects are denoted by \(u_i\), whereas
\(v_i\) denotes spatially dependent area level random effects, for which the authors
use an intrinsic conditional autoregressive (ICAR) prior. They explore the use of a
Gaussian process prior over the function \(\mu(\cdot)\) as well as a penalized spline
approach. For their Gaussian process prior, they assume a random walk of order
one. The multinomial model
\begin{equation}
  (n_{1i}, \ldots, n_{L_i i}) \sim \hbox{Multinomial} \left(n_i ; \frac{N_{1i} /
    w_{(1)i}}{\sum_{l=1}^{L_i} N_{li} / w_{(l)i}}, \ldots, \frac{N_{L_i i} /
    w_{(L_i)i}}{\sum_{l=1}^{L_i} N_{li} / w_{(l)i}} \right)
\end{equation}
is used for poststratification, where \(n_{li}\) and \(N_{li}\) represent the known
sample size and unknown population size respectively for poststrata cell $l$ in area
$i$. The cells are determined by the unique weights in area $i$, with the value of
the weight represented by $w_{(l)i}$. Although \citet{van16} implement their model
with a Bernoulli data example, this is a type of a Generalized Additive Model, and
thus other response types in the exponential family may be used as well.

\section{Likelihood-based inference using the sample distribution}\label{sec:
  samp_dist}

The pseudo-likelihood methods discussed in Section~\ref{sec: pseudo} require
specification of a superpopulation model, which is a distribution which holds for all
units in the finite population.  However, validating the superpopulation model based
on the observed sampled values is generally not possible, unless the sampling design
is not informative, in which case, the distribution for the sampled units is the same
as for the nonsampled units.  Under an informative sampling design, the model for the
population data does not hold for the sampling data.  This can be seen by application
of Bayes' Theorem.  Suppose the finite population values \(y_{ij}\) are independent
realizations from a population with density \(f_p ( \cdot \mid \boldsymbol{x}_{ij},
\boldsymbol{\theta} )\), conditional on a vector of covariates
\(\boldsymbol{x}_{ij}\), and model parameters \(\boldsymbol{\theta}\).  Given
knowledge of this superpopulation model, as well as the distribution of the inclusion
variables, the distribution of the sampled values can be derived.  Define the sample
density, \(f_s\), \citep{pfe98a} as the density function of \(y_{ij}\), given that
\(y_{ij}\) has been sampled, that is,
\begin{equation}\label{E: sampDist}
  f_s (y_{ij} \mid \boldsymbol{x}_{ij}, \boldsymbol{\theta}) = f_p (y_{ij} \mid
    \boldsymbol{x}_{ij}, \boldsymbol{\theta}, I_{ij} = 1) = \frac{P ( I_{ij} = 1 \mid
    y_{ij}, \boldsymbol{x}_{ij}, \boldsymbol{\theta}) f_p (y_{ij} \mid
    \boldsymbol{x}_{ij}, \boldsymbol{\theta})}{P (I_{ij} = 1 \mid
    \boldsymbol{x}_{ij}, \boldsymbol{\theta})}.
\end{equation}
From \eqref{E: sampDist}, the sample distribution differs from the population
distribution, unless \(P(I_{ij} = 1 \mid y_{ij}, \boldsymbol{x}_{ij}) = P(I_{ij} = 1
\mid \boldsymbol{x}_{ij})\), which occurs in ignorable sampling designs.  Note that
the inclusion probabilities, \(\pi_{ij}\), may differ from the probabilities \(P (
I_{ij} = 1 \mid \boldsymbol{x}_{ij}, y_{ij}, \boldsymbol{\theta})\) in \eqref{E:
  sampDist}, because the latter are not conditional on the design variables
\(\boldsymbol{Z}\).

Equation \eqref{E: sampDist} can be used for likelihood-based inference if the
simplifying assumption that the sampled values are independent is made.  While this
is not true in general, asymptotic results given in \citet{pfe98a} justify an
assumption of independence of the data for certain sampling schemes when the overall
sample size is large.  However, direct use of \eqref{E: sampDist} for finite
population inference requires additional model specifications for the sample
inclusion variables \(P(I_{ij} = 1 \mid \boldsymbol{x}_{ij}, y_{ij})\) as well as
\(P(I_{ij} = 1\mid \boldsymbol{x}_{ij})\).  It was shown in \citet{pfe98a} that
\(P(I_{ij} = 1 \mid \boldsymbol{x}_{ij}, y_{ij}) = E_P ( \pi_{ij} \mid
\boldsymbol{x}_{ij}, y_{ij})\), and that \(P(I_{ij} = 1 \mid \boldsymbol{x}_{ij}) =
E_p ( \pi_{ij} \mid \boldsymbol{x}_{ij} )\), so that a superpopulation model still
needs to be specified for likelihood-based inference.

Ideally, one would like to specify a model for the sampled data, and to use this
model fit to the sampled data to infer the nonsampled values, without specifying a
superpopulation model.  \citet{pfe99} derived an important identity linking the
moments of the sample and population-based moments, which allows for likelihood
inference using the observed data, without explicit specification of a population
model.  They showed that
\begin{equation*}
  P (I_{ij} = 1 \mid y_{ij}, \boldsymbol{x}_{ij}) = E_p (\pi_{ij} \mid y_{ij},
    \boldsymbol{x}_{ij}) = 1/E_s(w_{ij} \mid y_{ij}, \boldsymbol{x}_{ij}).
\end{equation*}
Similarly, it was shown that
\begin{equation*}
  P (I_{ij} = 1 \mid \boldsymbol{x}_{ij} ) = E_p (\pi_{ij} \mid \boldsymbol{x}_{ij})
    = 1 / E_s (w_{ij} \mid \boldsymbol{x}_{ij}).
\end{equation*}
Combining these results with an application of Bayes' Theorem, as was done to arrive
at equation \eqref{E: sampDist}, gives the distribution for the nonsampled units in
the finite population
\begin{equation}\label{E: nonSampDist}
  f_c (y_{ij} \mid \boldsymbol{x}_{ij}) \equiv f_p (y_{ij} \mid \boldsymbol{x}_{ij},
    I_{ij} = 0) = \frac{E_s(w_{ij} - 1 \mid y_{ij}, \boldsymbol{x}_{ij}) f_s (y_{ij}
    \mid \boldsymbol{x}_{ij})}{E_s(w_{ij} - 1 \mid \boldsymbol{x}_{ij})},
\end{equation}
where \( f_c \) represents the density function of \(y_{ij}\), given that \(y_{ij}\)
has not been sampled.  This result allows one to specify only a distribution for the
sampled responses and a distribution for the sampled survey weights for inference on
the nonsampled units, without any hypothetical distribution for the finite
population.  Importantly, this allows for identification of the finite population
generating distribution \(f_p\) through the sample-based likelihood.  It also
establishes the relationship between the moments of the sample distribution and the
population distribution, allowing for prediction of nonsampled units.

In the small area estimation context, the goal is prediction of the small area means,
\(\bar{y}_i\), which requires estimation of \(E_p(y_{ij} \mid D_S)\) in \eqref{E: bp}
for the nonsampled units in each area \(i\).  Suppose there is an area-specific
random effect, \(v_i \overset{i.i.d.}{\sim} \phi (v)\), common to all units in the
population in small area \(i\), so that the population distribution can be written
\(f_p (y_{ij} \mid \boldsymbol{x}_{ij}, v_i, \boldsymbol{\theta})\).  \citet{pfe07}
used the result in \eqref{E: nonSampDist}, to show how small area means can be
predicted using the observed unit level data under an informative survey design.
Under the assumption that \(E_c ( y_{ij} \mid D_s, v_i ) = E_c ( y_{ij} \mid
\boldsymbol{x}_{ij}, v_i )\),
\begin{equation*}
  E_p ( y_{ij} \mid D_s, I_{ij} = 0 ) = E_c ( y_{ij} \mid D_s ) = E_c ( E_c ( y_{ij}
    \mid \boldsymbol{x}_{ij}, v_i ) \mid D_s ).
\end{equation*}
Combining this with \eqref{E: nonSampDist} allows for prediction of the small area
means after specification of a model for the sampled responses, \(f_s (y_{ij} \mid
\boldsymbol{x}_{ij}, v_i)\), and a model for the sampled weights, \(f_s (w_{ij} \mid
y_{ij}, \boldsymbol{x}_{ij}, v_i)\).

The model for the survey weights can be specified conditionally on the response
variables to account for the informativeness of the survey design.  Possible models
for the sample weights considered in the literature include the linear model
\citep{bea08}
\begin{equation*}
  w_{ij} = a_0 + a_1 y_{ij} + a_2 y^2_{ij} + \boldsymbol{x}^T_{ij}
    \boldsymbol{\alpha} + \epsilon_{ij},
\end{equation*}
and the exponential model for the mean \citep{pfe98a, kim02, bea08}
\begin{equation}\label{E: expWt}
  E_s (w_{ij} \mid \boldsymbol{x}_{ij}, y_{ij}) = k_i \exp \left(a y_{ij} +
    \boldsymbol{x}^T_{ij} \boldsymbol{\beta} \right).
\end{equation}
\citet{pfe07} considered the case of continuous response variables, \(y_{ij}\), and
modeled the sampled response data using the nested error regression model \eqref{E:
  bhf}.  The exponential model for the survey weights in \eqref{E: expWt} was used to
model the informative survey design.  Under this modeling framework, they showed that
the best predictor of \(\bar{Y}_i\) is approximately
\begin{equation}\label{E: pfeBP}
  E_p ( \bar{Y}_i \mid D_s ) = N^{-1}_i \left[ (N_i - n_i) \hat{\theta}_i + n_i
    \left\{ \bar{y}_i + \left( \bar{\boldsymbol{X}}_i - \bar{\boldsymbol{x}}_i
    \right)^T \boldsymbol{\beta} \right\} + (N_i - n_i) b \sigma^2_e \right],
\end{equation}
where \(\hat{\theta} = \hat{u}_i + \bar{\boldsymbol{X}}^T_i \boldsymbol{\beta}\).
The term \((N_i - n_i) b \sigma^2_e\) in \eqref{E: pfeBP} is an additional term from
the usual best predictor in the nested error regression model \eqref{E: bhf}, which
gives a bias correction proportional to the sampling error variance \(\sigma^2_e\).

\citet{novelo17} take a fully Bayesian approach by specifying a population level
model for the response, \(f_p (y_{ij} | \boldsymbol{x}_{ij}, \boldsymbol{\theta}) \),
as well as a population level model for the inclusion probabilities, \(f_p (\pi_{ij}
| y_{ij}, \boldsymbol{x}_{ij}, \boldsymbol{\theta}) \). Through a Bayes rule argument
similar to (\ref{E: sampDist}), they show that the implied joint distribution for the
sampled units is
\begin{equation}\label{E: novJD}
  \begin{split}
    f_s(y_{ij}, \pi_{ij} | \boldsymbol{x}_{ij}, \boldsymbol{\theta}) & = f_p(y_{ij},
      \pi_{ij} | \boldsymbol{x}_{ij}, \boldsymbol{\theta}, I_{ij} = 1) \\ 
    & = \frac{\pi_{ij} f_p (\pi_{ij} | y_{ij}, \boldsymbol{x}_{ij},
      \boldsymbol{\theta})}{E_{y_{ij} | \boldsymbol{x}_{ij},
      \boldsymbol{\theta}} \{E( \pi_{ij} | y_{ij}, \boldsymbol{x}_{ij},
      \boldsymbol{\theta}) \} } \times f_p (y_{ij} | \boldsymbol{x}_{ij},
      \boldsymbol{\theta}).
  \end{split}
\end{equation}
This joint likelihood for the sample can then be used in a Bayesian model by placing
a prior distribution on \(\boldsymbol{\theta} \). Note that \(\boldsymbol{x}_{ij}\)
can be split into two vectors corresponding to \(f_p (y_{ij} | \boldsymbol{x}_{ij},
\boldsymbol{\theta}) \) and \(f_p (\pi_{ij} | y_{ij}, \boldsymbol{x}_{ij},
\boldsymbol{\theta}) \) if desired.  Consequently, the covariates for the response
model and the inclusion probability model need not be the same.

Two computational concerns arise when using the likelihood as in (\ref{E: novJD}).
The first issue is that in general, the structure will not lead to conjugate full
conditional distributions. To this effect, the authors recommend using the
probabilistic programming language Stan (\citet{carpenter217}), which implements HMC
for efficient mixing. The second concern is that the integral involved in the
expectation term of (\ref{E: novJD}) needs to be solved for every sampled observation
at every iteration of the sampler.  If the integral is intractable, it will need to
be evaluated numerically, greatly increasing the necessary computation time.  They
show that if the lognormal distribution is used for the population inclusion
probability model, then a closed form can be found for the expectation. Specifically,
let \(f_p (\pi_{ij} | y_{ij}, \boldsymbol{x}_{ij}, \boldsymbol{\theta}) = f(\log \,
\pi_{ij} | \mu=y_{ij} \kappa + t(\boldsymbol{x}_{ij}, \boldsymbol{\theta}),
\sigma^2=\sigma^2_{\pi}) \), where $f(\cdot|\mu,\sigma^2)$ represents a normal
distribution with mean $\mu$ and variance $\sigma^2$, \(\kappa\) is a regression
coefficient, and \(t(\cdot)\) some function. Then
\begin{equation}\label{E: novJD2}
  f_s(y_{ij}, \pi_{ij} | \boldsymbol{x}_{ij}, \boldsymbol{\theta}) =
    \frac{f(\log \, \pi_{ij} |\mu = y_{ij} \kappa + t(\boldsymbol{x}_{ij},
    \boldsymbol{\theta}), \sigma^2=\sigma^2_{\pi})}{ \exp\{t(\boldsymbol{x}_{ij},
    \boldsymbol{\theta}) + \sigma^2_{\pi}/2 \} E_{y_{ij} | \boldsymbol{x}_{ij},
    \boldsymbol{\theta}} \ \{\exp(y_{ij} \kappa)\} } \times f_p (y_{ij} |
    \boldsymbol{x}_{ij}, \boldsymbol{\theta}).
\end{equation}
In other words, the moment generating function of the population response model can
be used to find the analytical form of the expression, as long as the moment
generating function is defined on the real line. This includes important cases such
as the Gaussian, Bernoulli, and Poisson distributions, which are commonly used in the
context of survey data.

\section{Binomial likelihood special cases}\label{sec: bin}

The special case of binary responses is of particular interest to survey
statisticians, as many surveys focus on the collection of data corresponding to
characteristics of sampled individuals, with a goal of estimating the population
proportion or count in a small area for a particular characteristic.  In this
section, some techniques for modifying a working Bernoulli or binomial likelihood
using unit-level weights to account for an informative sampling design are discussed.

Suppose the responses \(y_{ij}\) are binary, and the goal is estimation of finite
population proportions in each of the small areas \(i = 1, \dots, m\),
\begin{equation*}
  p_i = \frac{1}{N_i} \sum_{j \in \mathcal{U}_i} y_{ij}.
\end{equation*}
The pseudo-likelihood methods discussed in Section \ref{sec: pseudo} can be directly
applied to construct a working likelihood of independent Bernoulli distributions for
the sampled survey responses as in Equation \eqref{E: pseudoLikelihood}.
\citet{zha14} used these ideas to fit a survey-weighted logistic regression model,
with random effects included at both the county level and the state level, using the
GLIMMIX procedure within SAS, to estimate chronic obstructive pulmonary disease by age
race and sex categories within United States counties.  Another example can be found
in \citet{con10}, who used a Bernoulli pseudo-likelihood to estimate diabetes
prevalence within U. S. states by demographic groups.  Their model formulation was
similar to that used by \citet{zha14}, but they included an additional random effect
to account for spatial correlation.

\citet{mal99} proposed a method which is similar in spirit to the pseudo-likelihood
method, which uses the survey weights to modify the shape of the binomial likelihood
function.  Suppose there are \(D\) demographic groups of interest and let
\(\mathcal{S}_d\) be the sampled individuals belonging to demographic group \(d = 1,
\dots, D\).  Instead of the usual independent binomial likelihood \(\prod_{i d}
p^{m_{id}}_{id} ( 1 - p_{id})^{n_{id} - m_{id}}\), \citet{mal99} proposed a
sample-adjusted likelihood
\begin{equation}\label{E: malec}
  \prod_{i d} \frac{p^{m_{id}}_{id} (1 - p_{id})^{n_{id} -
    m_{id}}}{(p_{id}/\bar{w}_{1d} + (1 - p_{id})/\bar{w}_{0d})^{n_{id}}},
\end{equation}
where
\begin{equation*}
  \bar{w}_{1d} = \sum_{(i, j) \in \mathcal{S}_d} w_{ijd} y_{ijd} / \sum_{(i, j) \in
    \mathcal{S}_d} y_{ijd}
\end{equation*}
and
\begin{equation*}
  \bar{w}_{0d} = \sum_{(i, j) \in \mathcal{S}_d} w_{ijd} (1 - y_{ijd}) / \sum_{(i, j)
    \in \mathcal{S}_d} (1 - y_{ijd}).
\end{equation*}
The quantities \(\bar{w}_{1d}\) and \(\bar{w}_{0d}\) are used to represent sampling
weights for a demographic group \(d\) averaged over all individuals with and without
a characteristic of interest, respectively.  The justification of the denominator of
\eqref{E: malec} as an adjustment to the likelihood to account for informative
sampling is presented in \citet{mal99} through use of Bayes' rule and by considering
the empirical distribution of the inclusion probabilities.

An alternative approach to the pseudo-likelihood method is to attempt to construct a
new, approximate likelihood with independent components, which matches the
information contained in the survey sample.  Let
\begin{equation*}
  \hat{p}_i = \frac{\sum_{j \in \mathcal{S}_i} w_{ij} y_{ij}}{\sum_{j \in
    \mathcal{S}_i} w_{ij}},
\end{equation*}
be the direct estimate of \(p_i\) and let \(\hat{V}_i\) be the estimated variances of
\(\hat{p}_i\). Under a simple random sampling design, the variance of the direct
estimate \(\hat{p}_i\) is \(V_{SRS} (\hat{p}_i) = p_i (1 - p_i)/n_i\), which can be
estimated by \(\hat{V}_{SRS} (\hat{p}_i) = \hat{p}_i (1 - \hat{p}_i)/n_i\).  In
complex sampling designs, elements that belong to a common cluster or area may be
correlated.  Because of this, the information in the sample from a complex survey is
not equivalent to the information in a simple random sample of the same size.  The
design effect for \(\hat{p}_i\) is the ratio
\begin{equation*}
  d_i = d_i (\hat{p}_i) = \frac{\hat{V}_D (\hat{p}_i)}{\hat{V}_{SRS} (\hat{p}_i)} =
    \frac{n_i \hat{V}_D (\hat{p}_i)}{\hat{p}_i (1 - \hat{p}_i)},
\end{equation*}
and is a measure of the extent to which the variability under the survey design
differs from the variability that would be expected under simple random sampling.

The effective sample size, \(n'_i\), is defined as the ratio of the sample size to
the design effect
\begin{equation*}
  n'_i = \frac{n_i}{d_i} = \frac{p_i (1 - p_i)}{V_{SRS} (\hat{p}_i)}.
\end{equation*}
The effective sample size is an estimate of the sample size required under a
noninformative simple random sampling scheme to achieve the same precision to that
observed under the complex sampling design.  Typically, the effective sample size
\(n'_i\) will be less than \(n_i\) for complex sample designs.

Often the design effect is not available, either due to lack of available information
with which to compute it, or due to computational complexity.  In such cases, design
weights can be used for estimation of the effective sample size.  A simple estimate
of the effective sample size, which uses only the design weights was derived by
\cite{kis65}, and is given by
\begin{equation*}
  n'_i = \frac{(\sum_{j \in \mathcal{S}_i} w_{ij})^2}{\sum_{j \in \mathcal{S}_i}
    w^2_{ij}}.
\end{equation*}
Other estimates of the design effect which use the survey weights, sample sizes, and
population totals, and are appropriate for stratified sampling designs, can be found
in \citet{kis92}.

\citet{che14} and \citet{fra14} used the design effect and effective sample size to
define the `effective number of cases,' \(y^*_i = n'_i \hat{p}_i\).  The effective
number of cases, \(y^*_i\), were then modeled using a binomial, logit-normal
hierarchical structure.  The sample model for the effective number of cases is then
\begin{equation*}
  y^*_i \mid p_i \sim Binomial (n'_i, p_i), i = 1, \dots, m,
\end{equation*}
with a linking model of
\begin{equation*}
  \mbox{logit}(p_i) = \log \left(\frac{p_i}{1 - p_i}\right) = \boldsymbol{x}^T_i
    \boldsymbol{\beta} + v_i,
\end{equation*}
where the \(v_i\) are area-specific random effects.  Using the effective number of
cases and the effective sample size in a binomial model is an attempt to construct a
likelihood which is valid under a simple random sampling design, and will produce
approximately equivalent inferences as when using the exact, but possibly unknown or
computationally intractable likelihood.

Different distributional assumptions on the random effects can be made to accommodate
aspects of the data or different correlation structures particular to sampled
geographies.  Noting that it might be expected that areas which are close to each
other might share similarities, \citet{che14} decomposed the random effects \(v_i\)
into spatial and a non-spatial components, so that \(v_i = u_i + \varepsilon_i\),
where \(\varepsilon_i \overset{i.i.d.}{\sim} N(0, \sigma^2_\varepsilon)\), and
\begin{equation}\label{E: icar}
  u_i \mid u_j, j \in ne(i) \sim N (\bar{u}_i, \sigma^2_u/n_i).
\end{equation}
Here, \(ne(i)\) is the set of neighbors of area i and \(\bar{u}_i\) is the mean of the
neighboring spatial effects.  The spatial model in \eqref{E: icar} is known as the
intrinsic conditional autoregressive (ICAR) model \citep{bes74}.

\citet{fra14} introduced a time dependence structure into the random effect \(v_i\)
for situations in which there are data from multiple time periods available, and
applied their model to estimation of poverty rates using multiple years of American
Community Survey data.  In their formulation, the random effects have an AR(1)
correlation structure, so that the model becomes
\begin{equation*}
  \begin{split}
    y^*_{i, t} \mid p_{i, t} & \sim Binomial (n'_{i, t}, p_{i, t}), \ i = 1, \dots,
      m, \ t = 1, \dots, T \\
    \text{logit}(p_{i, t}) & = \boldsymbol{x}^T_{i, t} \boldsymbol{\beta}_t +
      \sigma^2_t v_{i, t} \\ v_{i, t} & = \phi v_{i, t - 1} + \varepsilon_{i, t}
  \end{split}
\end{equation*}
where \(|\phi| < 1\), and the \(\varepsilon_{i, t}\) are assumed to be i.i.d. \(N (0,
1 - \phi^2)\) random variables.  The unknown parameters \(\boldsymbol{\beta}_t\) and
\(\sigma^2_t\) are allowed to vary over time.  \citet{fra14} showed that the
reductions in prediction uncertainty can be meaningful when the autoregressive
parameter \(\phi\) is large, but that the reduction in prediction uncertainty is more
modest when \(|\phi| < 0.4\).  As noted by \citet{che14}, the inclusion of spatial or
spatio-temporal random effects has the added benefit that the dependent random
effects can serve as a surrogates for the variables responsible for dependency in the
data.

The above methods use the survey weights either to modify the shape of an independent
likelihood \citep{mal97, zhe03} to account for the informative design, or to estimate
a design effect in an attempt to match the information contained in the survey sample
to the information implied by an independent likelihood by adjusting the sample size
\citep{che14, fra14}.  Alternatively, one could specify a working independence model
for the sampled units and incorporate the survey design by using the survey
weights as predictors \citep{zhe03}, and to induce dependence through a latent
process model.

\section{Simulation study}\label{sec: sim}

Unit-level models offer several potential benefits (e.g., no need for benchmarking
and increased precision), however, accounting for the informative design is critical
at the unit-level. There are a variety of ways to approach this; however, the utility
of each approach is not apparent. We choose three methods that span different general
modeling approaches (pseudo-likelihood, nonparametric regression on the weights, and
differing sample/population likelihoods), in order to address this question. We
choose to sample a population based on existing survey data from a complicated design, and
make estimates for poverty (similar to SAIPE).

To construct a simulation study, we require a population for which the response is
known for every individual, in order to compare any estimates to the truth. It is
also desirable to have an informative sample. We treat the 2014 ACS sample from
Minnesota as our population (around 120,000 observations and 87 counties), and further sample 10,000 observations in order to
generate our estimates from the selected models. Ideally, we would mimic the survey
design used by ACS, however the design is highly complex which makes replication
difficult. Instead, we subsample the ACS sample with probability proportional to the reported sampling weights, \(w_{ij}^{(o)}\), using the Midzuno
method \citep{midz} from the \texttt{sampling}  package in \texttt{R} \citep{samp}. This results in a new
set of survey weights \(w_{ij}^{(n)}\), which are inversely proportional to the original
weights given in the ACS sample. Sampling in this manner results in a sample for
which the selection probabilities are proportional to the original sampling
weights. By comparing weighted and unweighted direct estimates, we show that sampling in this way yields an informative sample. We fit three models to the newly sampled dataset, and create county
level estimates of the proportion of the original ACS sample below the poverty level.

\paragraph{Model 1}
\begin{equation}
  \begin{split}
    y_{ij} | \mathbf{\boldsymbol{\beta,\mu}} & \propto
      \hbox{Bernoulli}(p_{ij})^{\stackrel{\sim}{w}_{ij}} \\
    \hbox{logit}(p_{ij})& = \boldsymbol{x}_{ij}' \boldsymbol{\beta} + u_{i}  \\
    u_i  & \stackrel{ind}{\sim} \hbox{N}(0, \sigma^2_{u}) \\
    \boldsymbol{\beta}   \sim \hbox{N}_p(\boldsymbol{0}_p, \boldsymbol{I}_{p \times p}
      \sigma^2_{\beta}) & ,  \quad \sigma_{u}   \sim \hbox{Cauchy}^+(0, \kappa_{u}),
  \end{split}
\end{equation}
where the weights \(\stackrel{\sim}{w}_{ij}\) are scaled to sum to the total
sample size, as recommended by \citet{sav16}. We incorporate a vague prior
distribution by setting $\sigma^2_{\beta}=10$ and $\kappa_{u}=5$. This approach is
based on the Bayesian pseudo-likelihood given in \citet{sav16}. The model structure
is similar to that of \citet{zha14}, although we use the psuedo-likelihood in a
Bayesian context rather than a frequentist one. Our design matrix \(\bm{X}\) includes
terms for age category, race category, and sex. We use poststratification by
generating the nonsampled population at every iteration of our MCMC, which we use to
produce our estimates based on \eqref{E: bp}. The poststratification cells consist of
the unique combinations of county, age category, race category, and sex, for which
the population sizes are known to us.

\paragraph{Model 2}
\begin{equation}\label{E: mod2}
  \begin{split}
    y_{ij} | \beta_0, f(w_{ij}), \bm{u,v} & \sim \hbox{Bernoulli}(p_{ij}) \\
    \hbox{logit}(p_{ij}) & = \beta_0 + f(w_{ij}) + u_i + v_i \\
    f(w_{ij}) | \gamma, \rho & \sim \hbox{GP}(0, \hbox{Cov}(f(w_{ij}), f(w_{i'j'})))
      \\
    \hbox{Cov}(f(w_{ij}), f(w_{i'j'})) & = \gamma^2\hbox{exp}\left(-\frac{(w_{ij} -
      w_{i'j'})^2 }{2\rho^2}\right)  \\
    \bm{u} | \tau, \alpha & \sim \hbox{N}(0, \tau \bm{D} (\bm{I} - \alpha
      \bm{W})^{-1}) \\
    v_i | \sigma^2_v & \sim \hbox{N}(0,\sigma^2_v), \quad i=1,\ldots,m  \\
    \beta_0 \sim \hbox{N}(0, \sigma^2_{\beta}), \quad \gamma &\sim \hbox{Cauchy}^+(0,
      \kappa_{\gamma}), \quad  \rho \sim \hbox{Cauchy}^+(0, \kappa_{\rho}) \\
    \tau \sim \hbox{Cauchy}^+(0, \kappa_{\tau}), \quad \alpha & \sim
      \hbox{Unif}(-1,1), \quad \sigma_v \sim \hbox{Cauchy}^+(0, \kappa_v),
  \end{split}
\end{equation}
where \(\bm{D}\) is a diagonal matrix containing the number of neighbors for each
area \(i=1,\ldots,m\) and \(\bm{W}\) is an area adjacency matrix. Again, we use a
vague prior distribution by setting $\sigma^2_{\beta}=10$ and
$\kappa_{\gamma}=\kappa_{\rho}=\kappa_{\tau}=\kappa_{v}=5$. This is similar to the
work of \citet{van16}, but using the squared exponential covariance kernel as in
\citet{si15}, rather than a random walk prior on \(f(\cdot)\). Additionally, we
choose to use the conditional autoregressive structure (CAR) rather than ICAR
structure on our random effects $\bm{u}$. Note that although \citet{van16} use the
weights scaled to sum to county sample sizes as inputs into the nonparametric
function \(f(\cdot)\), we attained better results by using the unscaled weights. We
use the multinomial model
\begin{equation}
  (n_{1k}, \ldots, n_{L_k k}) \sim \hbox{Multinomial} \left(n_k ; \frac{N_{1k} /
    w_{(1)k}}{\sum_{l=1}^{L_k} N_{lk} / w_{(l)k}}, \ldots, \frac{N_{L_k k} /
    w_{(L_k)k}}{\sum_{l=1}^{L_k} N_{lk} / w_{(l)k}} \right)
\end{equation}
to model the population weight values, in order to perform poststratification. In
this model, \(n_{lk}\) represents the sample size in poststrata cell $l$ in area $k$,
while \(N_{lk}\) represents the population size in the same cell. Poststratification
cells are determined by unique weight values within each county, denoted
\(w_{(l)k}\). Because all units in the same cell will share the same weight, by
determining the population size of each cell, the weights are implicitly determined,
and thus the population may be generated using the model specified in (\ref{E: mod2}).

\paragraph{Model 3}
\begin{equation}\label{E: mod3}
  \begin{split}
    y_{ij} \mid p_{ij} & \sim \hbox{Bernoulli}(p_{ij}) \\
    \text{logit} (p_{ij}) & = \boldsymbol{x}^T_{ij} \boldsymbol{\beta} + u_i \\
    \log (w_{ij}) \mid y_{ij} & \sim \boldsymbol{x}^T_{ij} \boldsymbol{\alpha} +
      y_{ij} * a + \epsilon_{ij} \\
    u_i & \overset{i.i.d.}{\sim} \hbox{N} \left( 0, \sigma^2_u \right) \\
    \epsilon_{ij} & \overset{i.i.d.}{\sim} \hbox{N} \left( 0, \sigma^2_\epsilon
      \right), \\
  \end{split}
\end{equation}
with vague \(\hbox{N}(0, 10)\) priors on the regression coefficients
\(\boldsymbol{\beta}, \boldsymbol{\alpha}\), and \(a\), and vague
\(\hbox{Cauchy}^+(0, 5)\) priors on the variance components \(\sigma_u\) and
\(\sigma_\epsilon\).  This model acts as a Bayesian extension of \citet{pfe07}.

All 3 models were fit via HMC using Stan \citep{carpenter217}.
We ran each model using two chains, each of length 2,000, and discarding the first
1,000 iterations as burn-in, thus using a total of 2,000 MCMC samples. Convergence was
assessed visually via traceplots of the sample chains, with no lack of convergence  detected. We repeated the simulation 50 times, with a sample size of 10,000 each
time. That is, we create 50 distinct subsamples from the ACS sample, and fit the
three models to each subsample. We compare the mean squared error (MSE), absolute
bias, 95\% credible interval coverage rate for county level estimates, and computation time in seconds for each model in Table~\ref{table:simSummary}. We also compare to a Horvitz-Thompson (HT) direct estimator as well as an unweighted mean (UW) direct estimator.

Each of the three model based estimators provides a substantial reduction in MSE
compared to the direct estimator, with Model 3 being the best in this
regard. Additionally, Model 1 gives a low bias, quite comparable to the direct estimate. Finally, we see that Model 1 requires substantially less computation time
compared to the other model-based estimators, especially when comparing to
Model~2. This suggests that if one wanted to scale the model to include more data,
such as estimates at a national level, Model~1 may be easier to work with. Computation times will vary depending on the specific resources used, however the main focus here is the relative time between models. Additionally, this simulation illustrates that it is feasible to fit Bayesian unit-level models in practice under reasonable computation times.

\begin{figure}[H]
  \centering
  \includegraphics[width=1.0\textwidth]{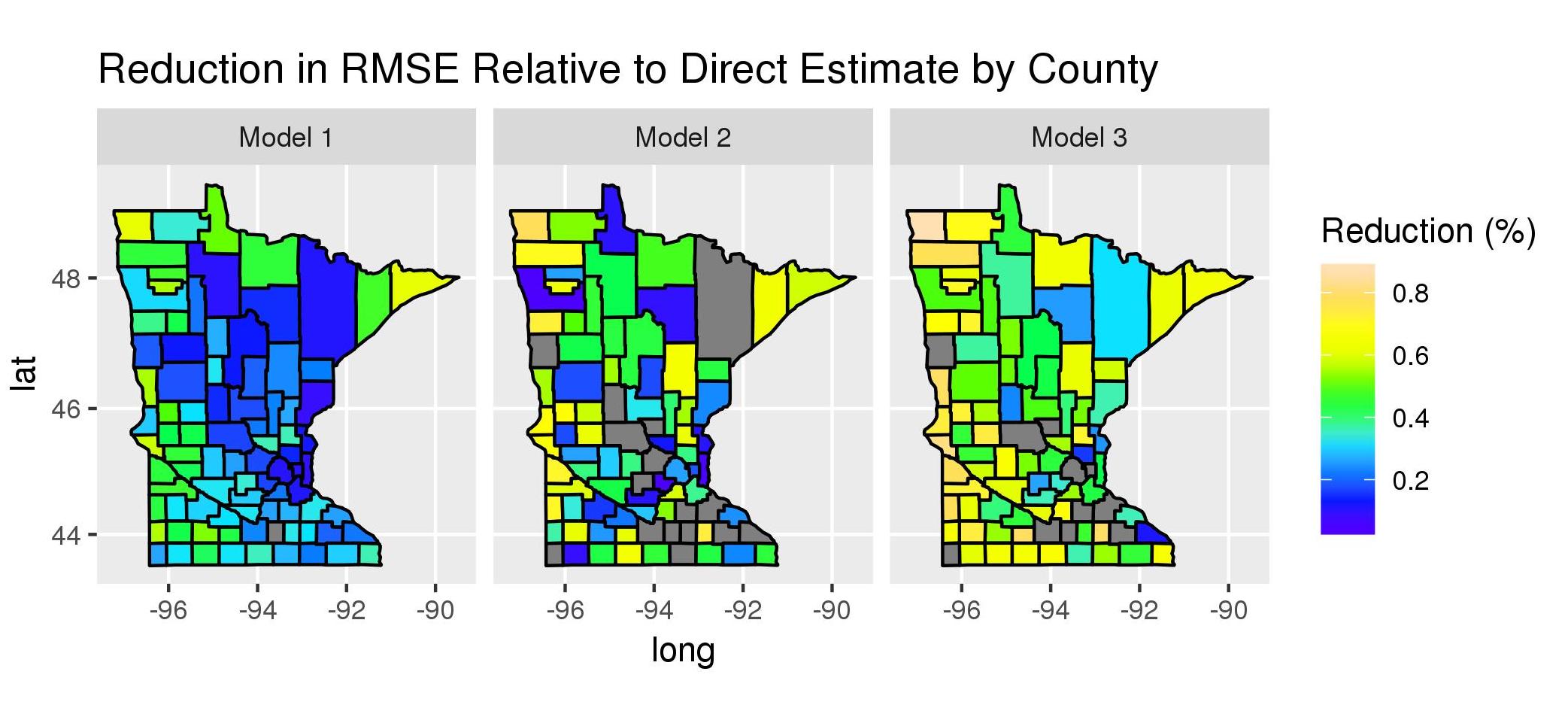}
    \caption{Model reduction in RMSE compared to the Direct estimates, averaged over 50 simulations. Counties that did not see a reduction are not plotted (shown in gray).}
  \label{fig:simRMSE}
\end{figure}

In Figure \ref{fig:simRMSE} we show the average reduction in RMSE, for each county, that was attained by the three model based estimators when compared to the HT direct estimator,
averaged over the 50 simulations. Counties that did not see a reduction are plotted in gray. There are some important differences between the model results here. Specifically, Model 1 achieves a reduction in nearly every county unlike the other two models, but Model 3 tends to achieve a greater reduction in RMSE in general when compared to Model 1. 

\begin{table}[t!]
  \begin{center}
  \begin{tabular}{c c c c c}
    \toprule
    Estimator & MSE & Abs Bias & CI Cov. Rate & Time  \\
    \midrule
    HT Direct    & 0.0044 & 0.0063 & 0.77 & NA  \\
    Model 1   & 0.0017 & 0.0089 & 0.86 & 107 \\
    Model 2   & 0.0017 & 0.0256 & 0.89 & 6627\\
    Model 3   & 0.0009 & 0.0172 & 0.94 & 407  \\
    UW Direct & 0.0050 & 0.0322 & 0.41 & NA \\
    \bottomrule
  \end{tabular}
  \caption{Simulation results: MSE, absolute bias, 95\% credible interval coverage rate, and computation time in seconds
    were averaged over 50 simulations in order to compare the direct estimator to
    three model based estimators and and unweighted direct estimate.}
  \label{table:simSummary}
  \end{center}
\end{table}

\section{Poverty Estimate Data Analysis}\label{sec: DA}

The Small Area Income and Poverty Estimates program (SAIPE) is a U.S. Census Bureau
program that produces estimates of median income and the number of people below the
poverty threshold for states, counties, and school districts, as well as for various
subgroups of the population. The SAIPE estimates are critical in order for the
Department of Education to allocate Title I funds.

The current model used to generate SAIPE poverty estimates is an area-level
Fay-Herriot model \citep{fay79} on the log scale. The response variable is the log
transformed HT direct estimates from the single year ACS of the number
of individuals in poverty at the county level.  The model includes a number of
powerful county level covariates such as the number of claimed exemptions from
federal tax return data, the number of people participating in the Supplemental
Nutrition Assistance Program (SNAP), and the number of Supplemental Security Income
(SSI) recipients. \citet{luery11} provides a comprehensive overview of the SAIPE
program, including the methodology used to produce various area-level estimates and
the covariates used in the model.

We use a single year of ACS data (2014 again) from Minnesota to fit the three models described in
Section~\ref{sec: sim}.  The model based estimators we present are not meant to
replace the current SAIPE methodology, but rather to illustrate how unit-level models
can be used in an informative sampling application such as this one. The model-based
predictions of the proportion of people below the poverty threshold by county under
each method are presented and compared with a direct estimator.

\begin{figure}[H]
  \centering
  \includegraphics[width=1.0\textwidth]{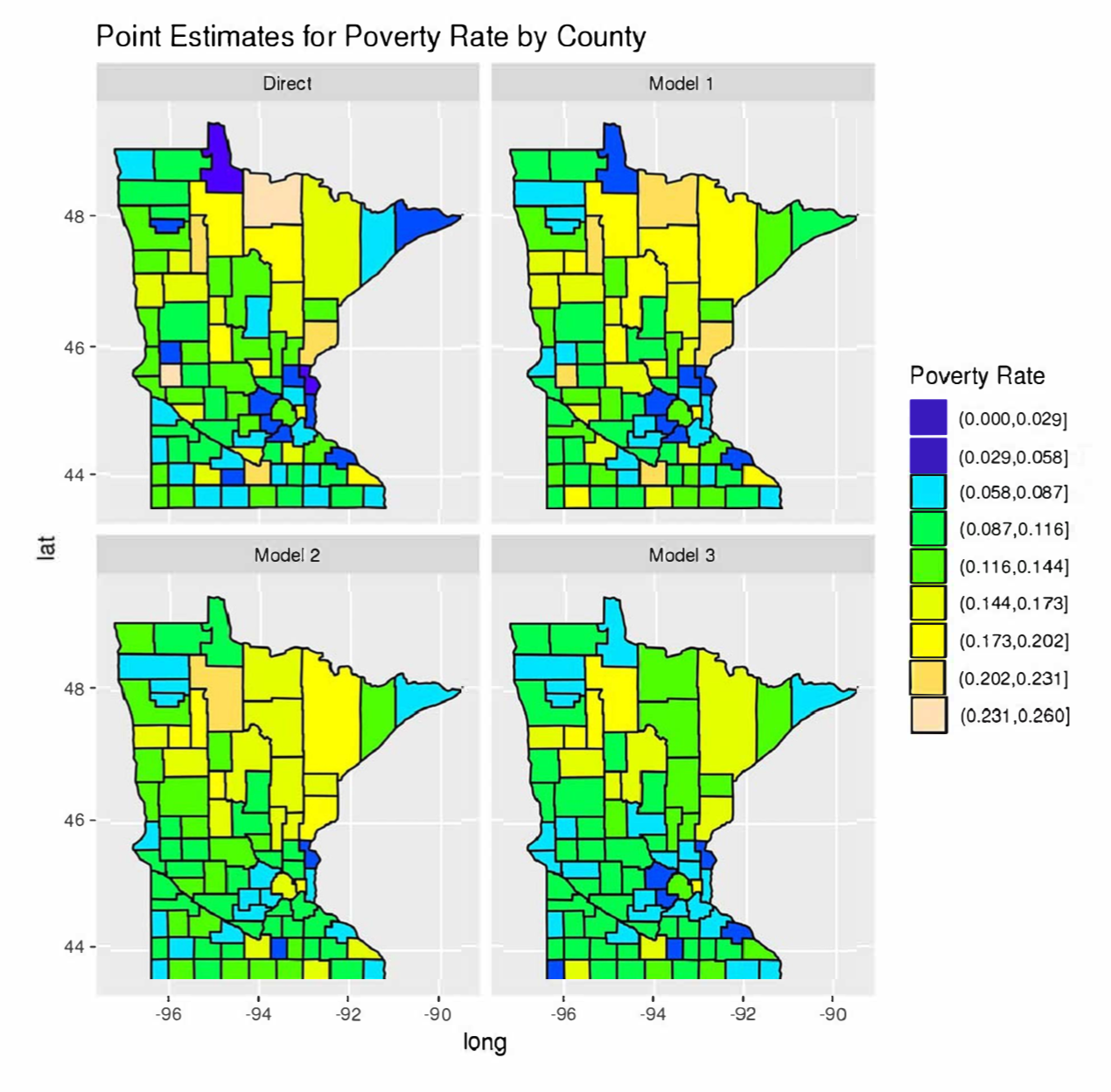}
    \caption{Noise infused HT direct and model based point estimates of poverty rate by county for Minnesota in 2014.}
  \label{fig:dataPoint}
\end{figure}

\begin{figure}[H]
  \centering
  \includegraphics[width=1.0\textwidth]{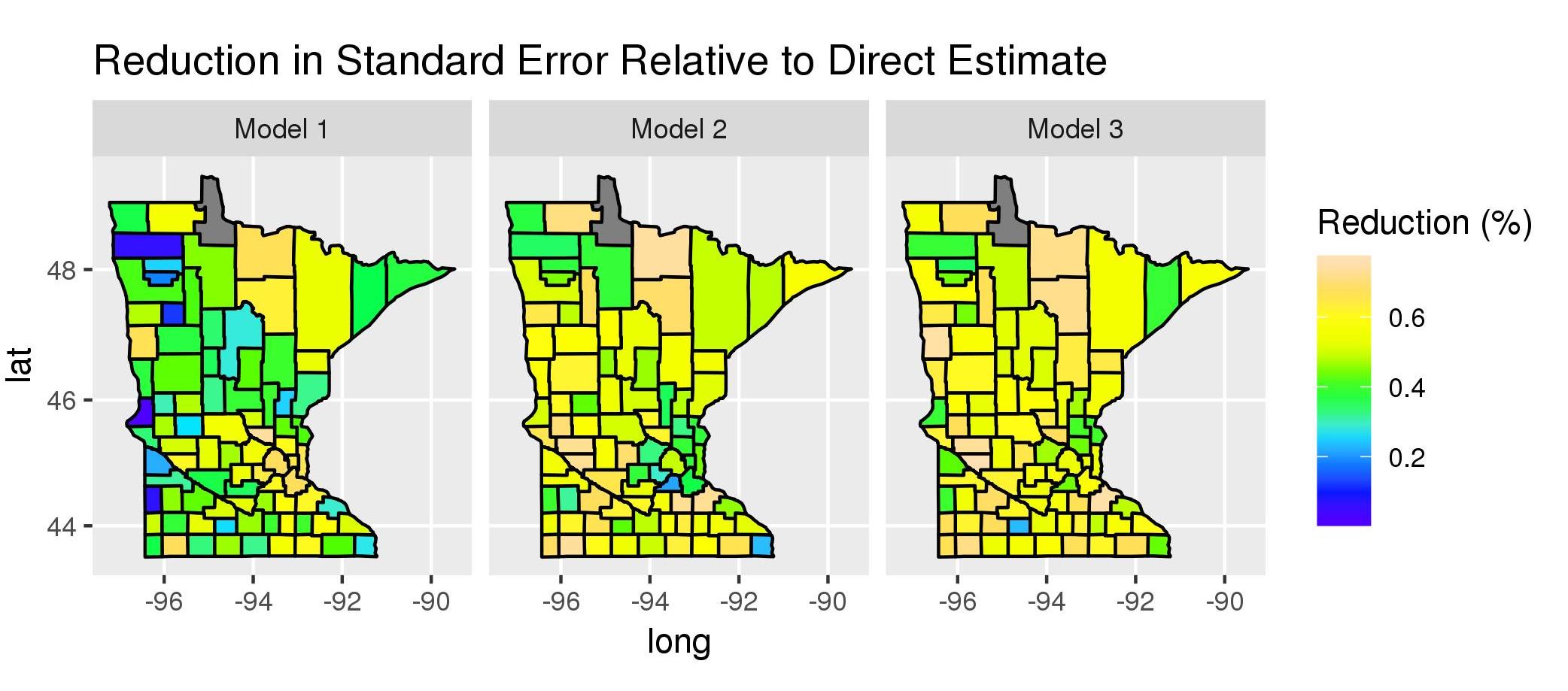}
    \caption{Model based reduction in standard errors for poverty rate by county. Counties that did not see a reduction are not plotted (shown in gray).}
  \label{fig:dataSE}
\end{figure}

In Figure~\ref{fig:dataPoint} we show the estimate of the proportion of people below
the poverty level by county for each of the model-based estimators as well as the
HT direct estimator. Note that a small amount of noise has been added to the HT direct estimates as a disclosure avoidance practice. All of the estimates here seem to capture the same general spatial
trend. The model based estimates resemble smoothed versions of the direct estimates, especially in the more rural areas of the state. Small sample sizes can lead to direct estimates with high variance, but the model based approaches can ``share information" across areas, which leads to more precise estimates. We also compare the reduction in model based standard errors when compared to the HT direct estimate in Figure \ref{fig:dataSE}. This illustrates the
precision that is gained by using a model-based estimator rather than a direct
estimator in a SAE setting. Model~3 in particular appears to have the lowest standard
errors in more rural areas and Model~1 seems to have lower standard errors in more populated areas. For this particular application, all three of the models we explored would be
valid choices, with substantial reductions in RMSE as shown in Section ~\ref{sec:
  sim}.

In this case, the population cell sizes were known, however in many applications they
may not be, in which case Model 2 would likely be the best option. In other cases
where incorporating covariate information is desired, Model 2 is not well equipped to
make estimates. This application was conducted for a single state, however if one
wanted to scale the analysis, for example making estimates for every county in the
United States, Model 1 appears to be the most computationally efficient. An approach
similar to Model 2, albeit using a different nonlinear regression approach from the
Gaussian Process regression considered here, may also be computationally
efficient. \citet{van16} reported strong results using splines for this
setup. Overall we found that each of these unit-level methods can offer precise
area-level estimates, however, the properties of the particular dataset under
consideration as well as the goals of the user should drive which model is selected.

Modeling poverty counts at the unit level has a number of benefits when compared to
area-level models. Specifically, the current SAIPE model is on the log scale, and
thus cannot naturally accommodate estimates for areas with a corresponding direct
estimate of zero, whereas unit-level modeling need not be on the log scale, and thus
does not suffer from this problem. Additionally, making predictions at multiple
spatial resolutions is straightforward in the unit-level setting, as predictions can
be generated for all units in the population and then aggregated as necessary, i.e.,
the so-called bottom-up approach. Under a unit-level approach, one could generate
poverty estimates at both a county level and school district level under the same
model.  In addition to these structural benefits, Table \ref{table:simSummary}
illustrates that unit-level models have the capacity to provide substantial
reductions in MSE and variance when compared to direct estimators.

\section{Conclusion}\label{sec: conc}

Through a comprehensive methodological review we have demonstrated that unit-level
models pose many advantages relative to area-level models. These advantages include
increased precision and straightforward spatial aggregation (the so-called
benchmarking problem), among others. Estimation of unit-level models requires
attention to the specific sampling design. That is, the unit response may be
dependent on the probability of selection, even after conditioning on the design
variables. In this sense, the sampling design is said to be \textit{informative} and
care must be taken in order to avoid bias.

In the context of small area estimation, we have described several strategies for
unit-level modeling under informative sampling designs and illustrated their
effectiveness relative to design-based estimators (direct estimates). Specifically,
our simulation study (Section~\ref{sec: sim}) illustrated three model-based
estimators that exhibited superior performance relative to the direct estimator in
terms of MSE, with Model~3 performing best in this regard. Among the three models
compared in this simulation, Model~1 displayed the lowest computation time relative
to the other model-based estimators and, therefore, may be advantageous in
higher-dimensional settings.

The models in Section~\ref{sec: sim} (and Section~\ref{sec: DA}) constitute modest
extensions to models currently in the literature. Specifically, Model~2 provides an
extension to \citet{van16}, whereas Model~3 can be seen as a Bayesian version of the
model proposed by \citet{pfe07}. With these tools at hand, there are many
opportunities for future research. For example, including administrative records into
the previous model formulations constitutes one area of active research as care needs
to be taken to probabilistically account for the record linkage. Methods for
disclosure avoidance in unit-level models also provides another avenue for future
research. In short, there are substantial opportunities for improving the models
presented herein. In doing so, the aim is to provide computationally efficient
estimates with improved precision. Ultimately, this will provide additional tools for
official statistical agencies, survey methodologists, and subject-matter scientists.

\section*{Acknowledgements}

This research was partially supported by the U.S. National Science Foundation (NSF) and the U.S. Census Bureau under NSF Grant SES-1132031, funded through the NSF-Census Research Network (NCRN) program, and NSF SES-1853096. Support for this research at the Missouri Research Data Center (MURDC), through the University of Missouri Population, Education and Health Center Doctoral Fellowship, as well as through the U.S. Census Bureau Dissertation Fellowship Program is also gratefully acknowledged. This article is released to inform interested parties of ongoing research and to encourage discussion. The views expressed on statistical issues are those of the authors and not those of the NSF or U.S. Census Bureau. The DRB approval number for this paper is CBDRB-FY19-506.

\baselineskip=14pt 
\bibliographystyle{jasa}
\bibliography{unit_level}

%
\clearpage\pagebreak\newpage

\end{document}